# Electron-cyclotron plasma startup in the GDT experiment

D V Yakovlev[1,2], A G Shalashov[1,3], E D Gospodchikov[1,3], A L Solomakhin[1,2], V Ya Savkin[1,2] and P A Bagryansky[1,2]

[1] Budker Institute of Nuclear Physics, Siberian Branch of Russian Academy of Sciences, Novosibirsk, Russia
[2] Novosibirsk State University, Novosibirsk, Russia
[3] Institute of Applied Physics of Russian Academy of Sciences, Nizhny Novgorod, Russia

E-mail: d.v.yakovlev@inp.nsk.su; ags@appl.sci-nnov.ru

**Abstract**
The paper reports on a new plasma startup scenario in the Gas Dynamic Trap (GDT) magnetic mirror device. The primary 5 MW neutral beam injection (NBI) plasma heating system fires into a sufficiently dense plasma target ("seed plasma"), which is commonly supplied by an arc plasma generator. In the reported experiments, a different approach to seed plasma generation is explored. One of the channels of the electron cyclotron resonance (ECR) heating system is used to ionize the neutral gas and build up the density of plasma to a level suitable for NBI capture. After a short transition (about 1 ms) the discharge becomes essentially similar to a standard one initiated by the plasma gun. The paper presents the discharge scenario and experimental data on the seed plasma evolution during ECR heating, along with the dependencies on incident microwave power, magnetic configuration and pressure of a neutral gas. The characteristics of consequent high-power NBI discharge are studied and differences to the conventional scenario are discussed. A theoretical model describing the ECR breakdown and the seed plasma accumulation in a large scale mirror trap is developed on the basis of the GDT experiment.

## 1. Introduction

In a toroidal magnetic plasma confinement device a non-inductive plasma startup usually requires application of wave plasma heating methods (see e.g. [1, 2]), however a possibility for direct injection of high-pressure plasma stream across the magnetic field lines has also been explored [3]. A linear device, on the other hand, encourages direct injection of plasma or beams of charged particles along the magnetic field lines, making the discharge startup much less of an issue. However, such injection necessitates the coupling of the magnetic flux between the plasma source and the confinement region, which is not always easily achieved. The second problem, which is important mainly for a short-pulse experiment, is the residual plasma and excess neutral gas from the plasma generator, which can undermine all the efforts to thermally insulate plasma from the end-wall. Finally, in a reactor-grade magnetic mirror a direct power conversion unit or additional plasma heating system would be certainly preferred over occasionally used plasma generator. Therefore, a technique not relying on direct plasma injection is a vital technology for a magnetic mirror that aims at fusion applications.

Plasma startup by waves of electron cyclotron frequency range was previously examined in virtually every major magnetic mirror installation. The notable examples are Phaedrus [4], Tara [5], TMX-U [6] and OGRA-4 [7] experiments. Physics of ECR breakdown with high-power millimeter wave radiation was studied in detail for the ECR source of multi-charged ions SMIS-37 [8]. The most recent study was performed in GAMMA-10 tandem mirror [9]. Central cell electron cyclotron resonance heating (ECRH) system was used for gas breakdown and plasma build-up, followed by primary heating by ICRF waves. It is concluded that ECR startup is similar to the standard startup with axial plasma guns, at least when it comes to plasma parameters after the initial transient stage. With ECRH power of 100kW the transition to stationary conditions is 10-15 ms longer than in the gun-discharge [10]. Interestingly enough, this experiment did not reveal any influence of wave polarization, microwave beam orientation and incident power on discharge parameters.

In this paper we investigate an ECR plasma startup scenario developed for the large-scale Gas Dynamic Trap (GDT) magnetic mirror experiment. GDT is an axially symmetric open-ended linear plasma confinement device with plasma flux suppression by high-field magnetic mirror coils [11]. Main plasma heating by 5 MW neutral beam injection (NBI) is aided by auxiliary 0.8 MW ECRH system [12-14]. The discharge in the GDT is routinely initiated by 1.5 MW washer-stack arc generator, which feeds plasma to the central cell through one of the magnetic mirrors. This plasma is successively used as a target for NBI, which eventually heats it up and establishes steady state density and temperature profiles.

To create a substitute for a gun plasma target, we focus a 400 kW / 54.5 GHz / X-polarized microwave beam at the fundamental EC resonance surface near the machine axis. After a prompt ionization of neutral gas, a steady accumulation of seed plasma is observed. At a certain point, the seed plasma becomes dense enough for the NBI to take over. Compared to most previous magnetic mirror experiments, where ECR seed plasma was to be treated by ICRF heating, in GDT such plasma target has to be dense enough for a self-sustained NB capture, which must overcome electron drag, charge exchange losses and cover the ionization expenses. Compared to a more relevant TMX-U experiment, where the ECR seed plasma was used as a target for NBI in a minimum-B cell, in GDT such plasma target has to be generated in a simple magnetic mirror field, thus making it potentially vulnerable to interchange instabilities.

Although theory of ECR heating in open magnetic configurations was discussed in a wide context, peculiarities of the ECR breakdown were most deeply understood basing on dedicated experiments at OGRA-4 and SMIS-37 [7, 8]. However, essential features of these experiments were longitudinal propagation of microwaves with respect to the confining magnetic field, and relatively compact plasma. Opposite to it, the GDT ECRH system uses inclined launching of the microwave beam, what in combination with the large mirror-to-mirror length results in a new regime of the breakdown. This motivates us to spent some more time for basic studies of the breakdown and the seed plasma accumulation phase resulting in a new theoretical model supported by the experiments.

The paper is organized as follows. Section 2 presents the general information about the GDT setup. Section 3 discusses the experiments on ECR-breakdown and describes the evolution of seed plasma sustained purely by ECRH. Discharges initiated by ECR-breakdown and supported by high-power NBI are discussed in section 4; in particular, the difference between the gun- and RF-initiated discharges in terms of general plasma confinement is discussed. A comprehensive theory of ECR startup in GDT is developed in section 5. A brief summary of the experiment is given in Conclusions. Finally, some of key formulas used for theoretical modeling are derived analytically in Appendix.

## 2. The gas dynamic trap device

GDT is a large-scale axially symmetric magnetic mirror device with 7-m-long central cell and two end-tanks, which house the expanding plasma flux (see figure 1). Ratio between maximum and minimum values of magnetic field (the on-axis mirror ratio) is $R = 30$. Design, diagnostics, physics of plasma confinement and main goals of this machine are described in [11, 15-21]. GDT parameters in the regime of operation used for the present study are listed in table 1.

The device is distinguished with fairly good confinement that allows reaching local plasma $\beta$ up to 60% over the ambient magnetic field of 0.7 T and $n_e T_e \approx 1.2 \times 10^{13} \text{cm}^{-3} \times 0.7 \text{keV}$, where $T_e$ and $n_e$ are the electron temperature and density, respectively. Plasma in GDT consists of two components with different mean energies. First one is the bulk plasma serving as a target for NBI. Due to high collisionality the bulk plasma is characterized by isotropic Maxwell velocity distribution with a temperature of $100 - 250$ eV (in a pure NBI discharge) both for ions and for electrons, therefore it is confined in a gas-dynamic regime. The second energy fraction is represented by fast ions with a mean energy of about 10 keV sustained by the NBI. The energy confinement time of fast ions is limited by the electron drag in bulk plasma, which results in anisotropic velocity distribution with relatively small angular spread. Corresponding space distribution of fast ions in the central cell has its characteristic density and pressure peaks near the reflection points at $R = 2$. The gas-dynamic outflow of bulk plasma is replenished by both the external gas feed and the influx of fast atoms. Plasma is constrained radially by two ring limiters with inner radii projecting to $r = 15$ cm at the midplane. The voltage applied between the limiters and the end-plates induces a sheared azimuthal flow in plasma [22], which keeps down the transverse energy loss induced by lower interchange modes excited in the GDT`s magnetic configuration.

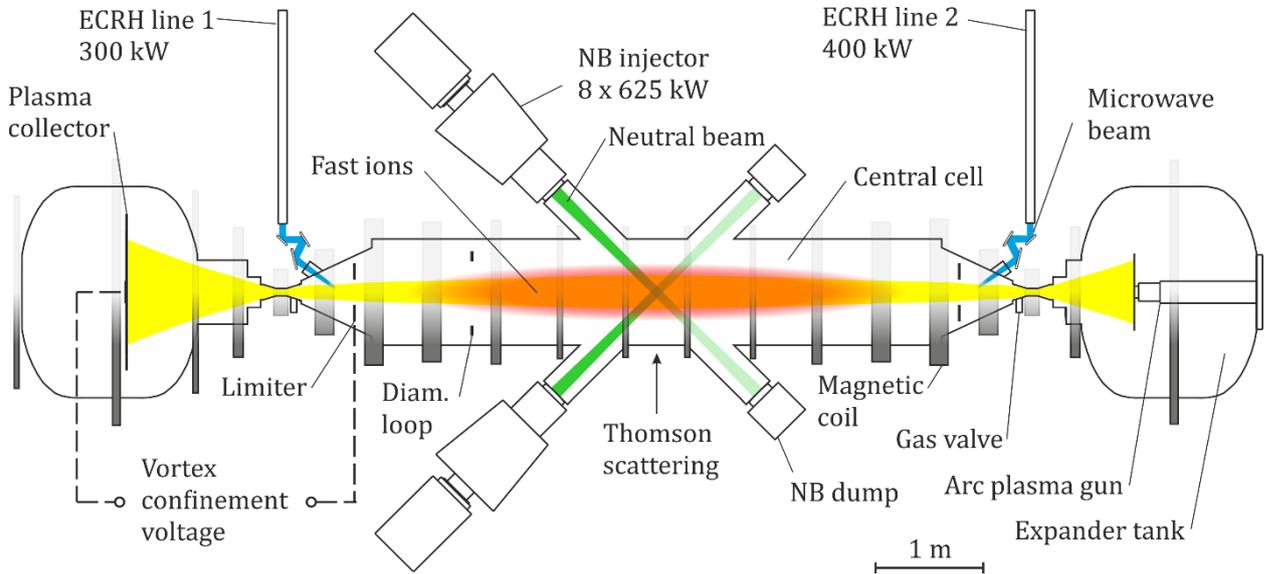

**Figure1.** Schematic of the GDT.

**Table 1.** GDT parameters used in EC startup experiment

| | |
|---|---|
| Mirror-to-mirror distance | 7 m |
| Plasma radius at midplane | 0.15 m |
| Magnetic field at midplane | 0.35 T |
| Mirror ratio | 30 |
| ECRH power | 0.4 MW |
| ECRH frequency | 54.5 GHz |
| Electron temperature | ~ 0.25 keV |
| Total neutral beam (NB) power | 4.5-5 MW |
| Trapped NB power | 1-3 MW |
| NB pulse duration | 4.5 ms |
| NB injection angle | 45° |
| Energy of NB particles | 23 keV |
| Mean energy of fast ions | ~10 keV |
| Plasma density at midplane | $0.8\text{-}1.3\times10^{19}\,\text{m}^{-3}$ |

Central cell gas feed system consists of two pulsed solenoid valves. Each valve is connected to a gas box collar attached to vacuum vessel near the plasma choke points (figure 1). The gas box opens up on plasma by a ring of azimuthally spread leak holes slightly inclined towards the machine center. The amount of injected gas is regulated both by valve inlet pressure and by pulse duration. The experiment is conducted with a deuterium plasma.

For additional plasma heating, GDT is equipped with ECRH system operating at 54.5 GHz with total power of up to 0.8MW [12-14]. Two separate microwave beams (figure 1) are launched obliquely into plasma near the magnetic mirrors from the high-field side with the extraordinary wave polarization. Along their way to the first harmonic EC resonance, the waves experience strong refraction, which is determined by the particular magnetic field distribution and plasma density profile. Oblique propagation of microwave beam in a highly refractive plasma leads to remarkably reach and rather unusual for a linear device ECRH physics that has been studied recently in [23-26].

## 3. ECR breakdown and seed plasma build-up

*3.1. ECR start-up scenario*

A typical scenario of ECR breakdown experiment is shown in figure 2. As in the standard regime with the plasma gun, central cell vessel is pre-filled with neutral $D_2$ through a 10 ms gas feed pulse. The valves are shut off at $t = 1$ ms. In most shots ECRH operates from $t = 1.5$ to $t = 4$ ms, while NBI heating operates from $t = 3.6$ to $t = 9$ ms. Note that discharges with NBI heating are discussed in section 4.

The evolution of average neutral $D_2$ density in the central cell (shown in figure 2) was calculated with *MolFlow+* software [27]. To obtain the neutral density distribution, a molecular gas flow simulation was performed with taking into account the realistic geometry of the GDT`s central cell, expander tanks and gas interface. The bulk of central cell pumping is undertaken by the vessel wall, which is recoated with titanium after each discharge. A corresponding sticking probability value of 0.05 for $D_2$ was selected based on recent measurements [28]. The real-life performance of each gas feed line was estimated as 20 mbar·l/s, leaving aside the effects of gas line capacity and influence of pulsed magnetic fields on gas feed valves. Overall, we estimate the error in resulting neutral density to be below 50%.

Each of the two available ECRH channels was found to be capable of ECR breakdown (including simultaneous operation). However, to simplify the analysis, only the 400 kW gyrotron (line 2 in figure 1) was selected for further experiments. Figure 3 shows magnetic configurations used in the experiments and corresponding variation of the magnetic field strength along central ray of the microwave beam propagating in neutral gas. For most shots, the magnetic field near the ECR region is adjusted such that the microwave beam intersects the first harmonic ECR surface approximately on the axis of the machine (figure 3, b). Note that the resonance is met twice along the beam – the second ECR surface is located in the launching port inside the vacuum chamber. If some cold and rarefied residual plasma is present here, this resonance is screened with the cut-off surface that may result in partial reflection of the gyrotron radiation propagating from the low magnetic field side. For that reason the auxiliary resonance in the launching port is considered a "parasitic" resonance.

By coincidence, the selected magnetic configuration is not different from the one used for the main plasma heating at the developed stage of a discharge (despite entirely different beam path in plasma). It allows, in principle, the discharges with ECR breakdown and additional ECR heating with the other available gyrotron during the developed stage of a discharge. However, for technical reasons such advanced scenario is not yet available. Other magnetic configurations in figure 3 will be explained further.

As it will be shown later, available microwave power far exceeds the breakdown threshold of a neutral gas at the fundamental harmonic. Indeed, the ECR breakdown is found in a very broad range of experimental conditions. Then, the remaining task for the ECR startup is to build up a plasma with sufficient density, which enables accumulation of fast ions during NBI. This process will be studied in the next section. In present one we focus on plasma dynamics at the early stage before the NBI switch-on.

As a typical example, figure 4 shows the line-averaged plasma density measured at the midplane by a dispersion interferometer (DI) [29] and stray radiation signal measured by a microwave antenna diode. According to stray radiation signal, after a short ~100-300 μs period of poor microwave absorption, plasma becomes opaque and stays such until the gyrotron switch-off. At the very moment of the breakdown and at least 100 μs later, the line density of plasma is below the detection limit of the DI diagnostic, which is $\sim 2 \cdot 10^{13}$ cm$^{-2}$. After that, the density grows linearly for 1-2 ms and then halts at a certain saturated value. When the microwave power is switched off and no NBI heating is applied, plasma decays within 3-5 ms.

Currently, the clearest picture of seed plasma generation is provided by the DI diagnostic. Attempts to measure plasma profiles with a triple probe were only partially successful: even at $r = 16$ cm electron temperature values reached ~100 eV and eventually damaged the probe.

The values of electron density and temperature yielded by Thomson scattering (TS) at the line density saturation stage exhibit large shot-to-shot fluctuations, caused by limited TS resolution and possible issues with a seed plasma stability. The only conclusive data provided by the TS diagnostic is that the local electron density measured at radial positions of 0, 3 and 6 cm does not exceed $\sim 3 \cdot 10^{12}$ cm$^{-3}$.

It is observed that standard shear stabilization leads to degradation of seed plasma. Biasing the limiters with 360 V at the flattop stage of line density triggers plasma decay within $\sim 2$ ms. With no other means to stabilize the interchange modes, one should expect substantial transverse transport during ECR plasma generation.

An insight into energy confinement time of ECR-generated plasma might be given based on diamagnetic flux measurements. Overall, the signals measured by diamagnetic loops are $\sim 2$ times weaker than usual noise of the diagnostic and are about 1-2% of those in a full power NBI discharge. Figure 5 shows average signals measured by $R = 1$ and $R = 2$ loops in a series of discharges with line density of $2 \cdot 10^{14}$ cm$^{-2}$. The difference between the diamagnetic fluxes, which is a difference in transverse plasma pressure, suggests an anisotropic distribution function of plasma components.

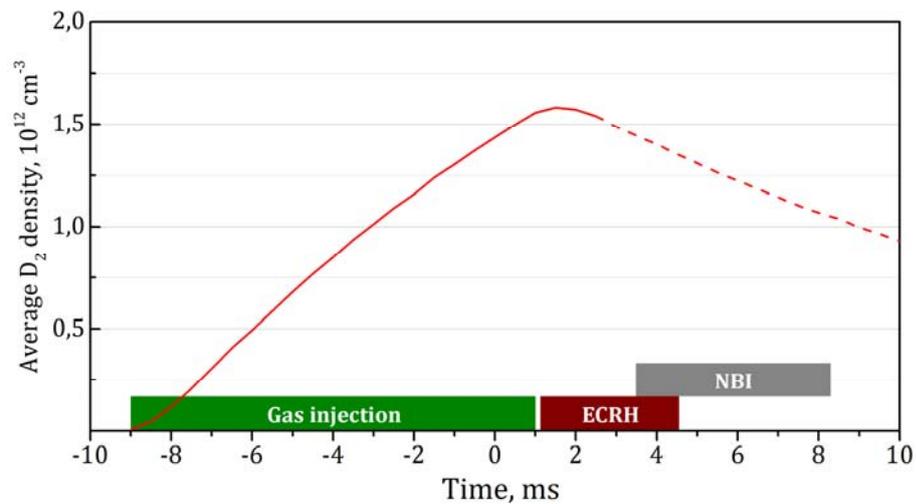

**Figure 2.** Scenario of ECR breakdown experiment and calculated average neutral $D_2$ density in the central cell.

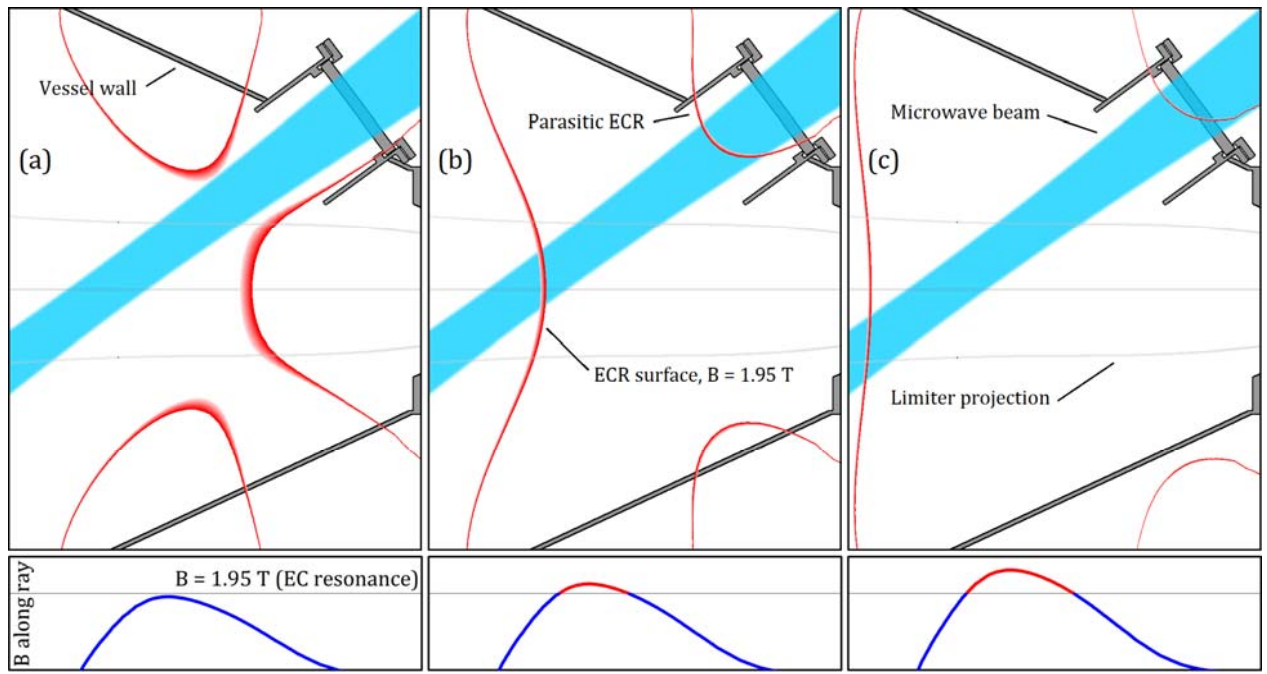

**Figure 3.** Magnetic configurations and microwave beam propagating in neutral gas (top) and variation of magnetic field strength along central ray of the beam (bottom): (a) – low field configuration without absorption of power, (b) – optimum configuration for ECR startup, (c) – high field configuration.

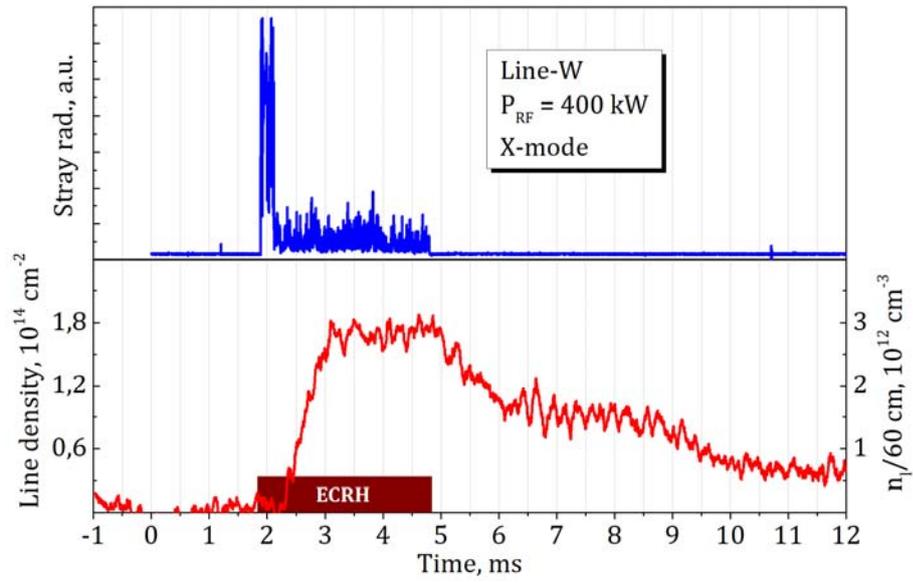

**Figure 4.** Stray microwave radiation (top) and midplane line-averaged density of plasma (bottom) in a pure ECRH discharge without NBI.

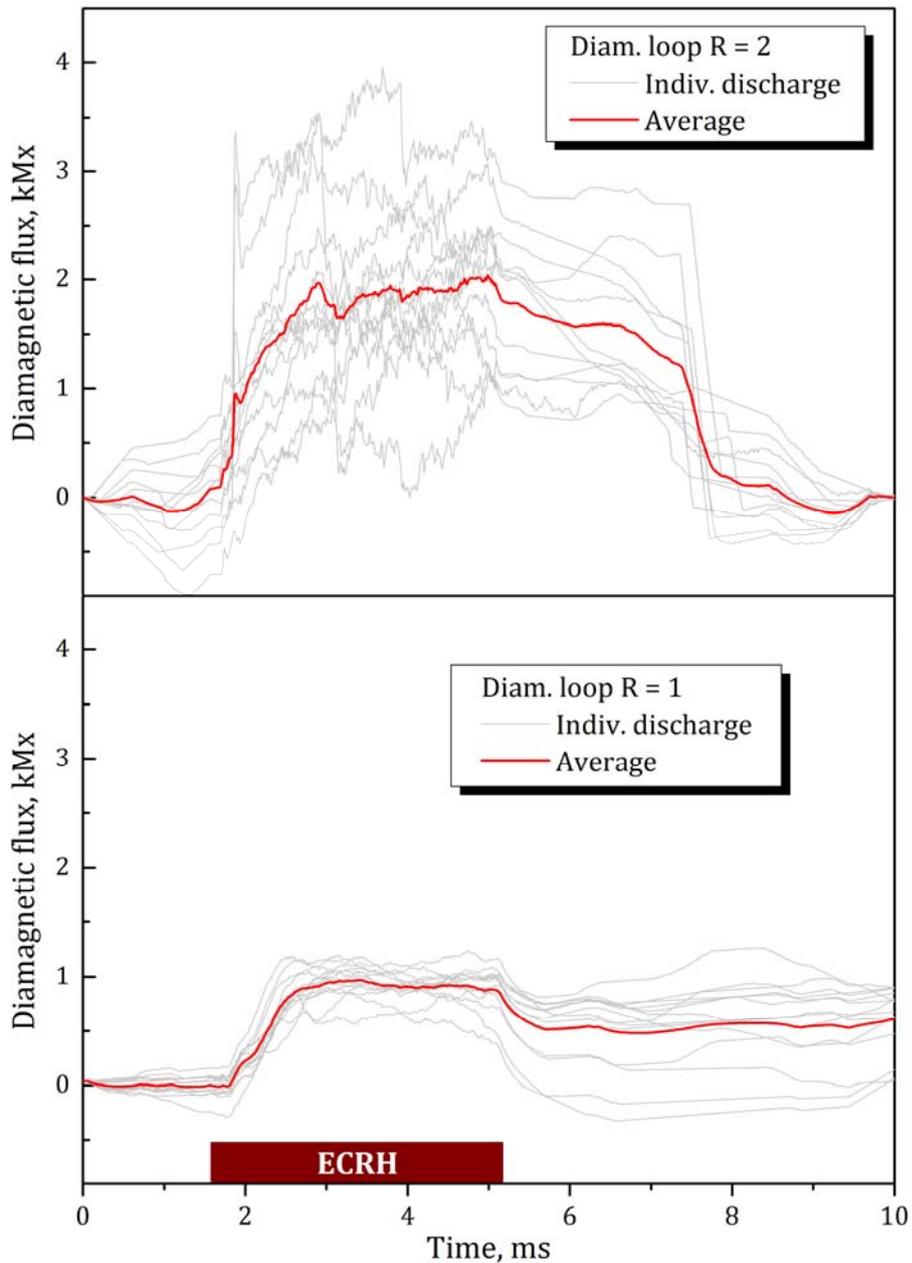

**Figure 5.** Averaged and raw diamagnetic fluxes of seed plasma. Top: $R = 2$ loop. Bottom: $R = 1$ loop (midplane).

*3.2. Experiments with variable ECR conditions*

A series of discharges with variable magnetic field was performed to determine the influence of ECR layer position on plasma build-up. Magnetic field was adjusted through variation of current in a coil located over the ECR port (figure 1). The rest of magnetic configuration was kept unmodified. Figure 3 illustrates the range of explored magnetic configurations. Note that the wave beams depicted in this figure do not take into account beam refraction by the seed plasma and are good only for a qualitative analysis of radial distributions of electrons affected by ECRH.

In case of excessively low field (figure 3, a) the microwave beam does not cross the resonance field directly, and so no accumulation of seed plasma is observed. This fact suggests that the first pass absorption at ECR layer is crucial to seed plasma generation and sustainment, as opposed to indirect ionization by scattered wave field in the vacuum vessel.

Shift of the ECR layer away from the magnetic mirror (figure 3, b-c) does not influence line plasma density or stray radiation signal. The difference, however, is observed in D-D neutron flux detector [30], used as an X-ray detector in this case (figure 6). Shift of the beam spot on the ECR layer from optimum position on axis to plasma periphery leads to generation of bremsstrahlung X-rays during the whole microwave pulse. The overheated electron population, which is almost certainly being generated during ECRH, is localized and in case of peripheral localization can directly reach parts of central cell vacuum vessel.

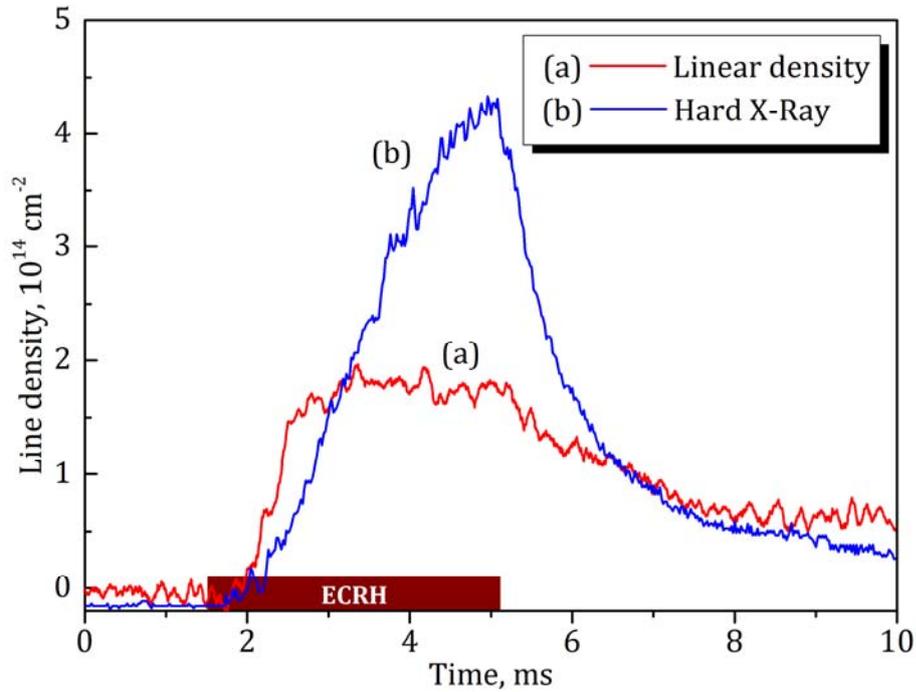

**Figure 6.** Hard X-ray signal and line-averaged density of plasma in a discharge with peripheral aiming of microwave beam (figure 3, c).

*3.3. Experiments with variable microwave power*

Figure 7 shows line density of seed plasma in a number of discharges with variable gyrotron output. It is observed that the line density is nearly identical for values of incident power > 150 kW. At lower values, the saturation level starts gradually decreasing, though the delay before the plasma build-up stays the same. Stray radiation signal is proportional to input microwave power up to 250 kW, and stays constant above this power level. Note that in this series the initial average density of neutral $D_2$ was $\sim 1.5 \cdot 10^{12}$ cm$^{-3}$.

All reported results have been obtained with microwave beam polarization, which would correspond to the extraordinary plasma wave. A series of discharges with the orthogonal polarization (the ordinary wave) has demonstrated much weaker plasma build-up. Full power in the ordinary mode was equivalent to about 10% of microwave power in the extraordinary mode. At power levels below 40 kW (10%) no accumulation of seed plasma is observed, while such threshold for the extraordinary mode is beyond the experimental accuracy. Note that the actual purity of O-mode is not precisely controlled and X-mode content may be up to 2%.

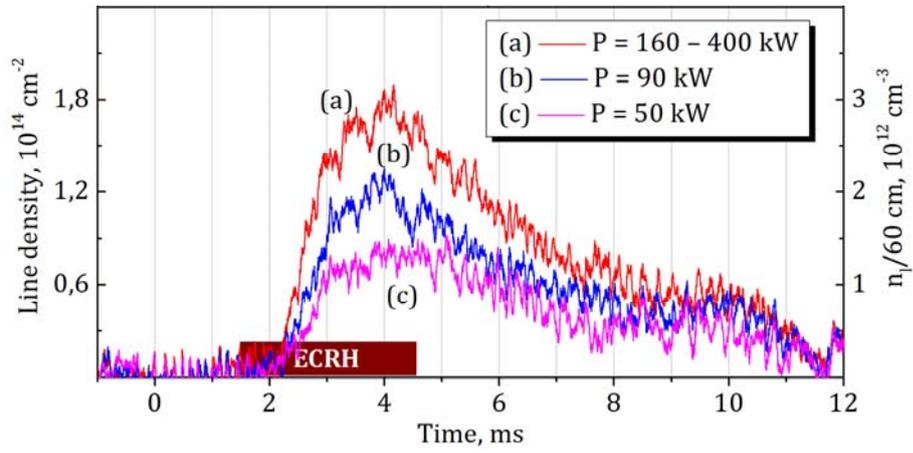

**Figure 7**. Line-averaged density of seed plasma in experiment with variable microwave power

*3.4. Experiments with variable pressure of source gas*

To determine the influence of initial density of neutral gas on ECR breakdown, a series of discharges with variable gas feed duration was carried out. The valves were shut off at $t = 1$ ms, but the opening trigger was shifted allowing to scan through gas feed durations from $\Delta t = 4$ ms to $\Delta t = 40$ ms. However, the resulting neutral density cannot be directly extrapolated from the gas feed duration, as the central cell pump-out time is 10-20 ms. To obtain the density distribution, a calculation similar to the one described in section 3.1 was performed.

Saturated level of line plasma density (figure 8) displays monotonic dependence on initial neutral gas pressure. The highest achieved line density equaled to $\sim 2.5 \cdot 10^{14}$ cm$^{-2}$. Probably even higher values of line plasma density attained with $\Delta t > 30$ ms could not be reached due to air breakdown in the ECRH line, likely caused by unacceptable level of reflected radiation from the parasitic ECR layer (figure 3, b).

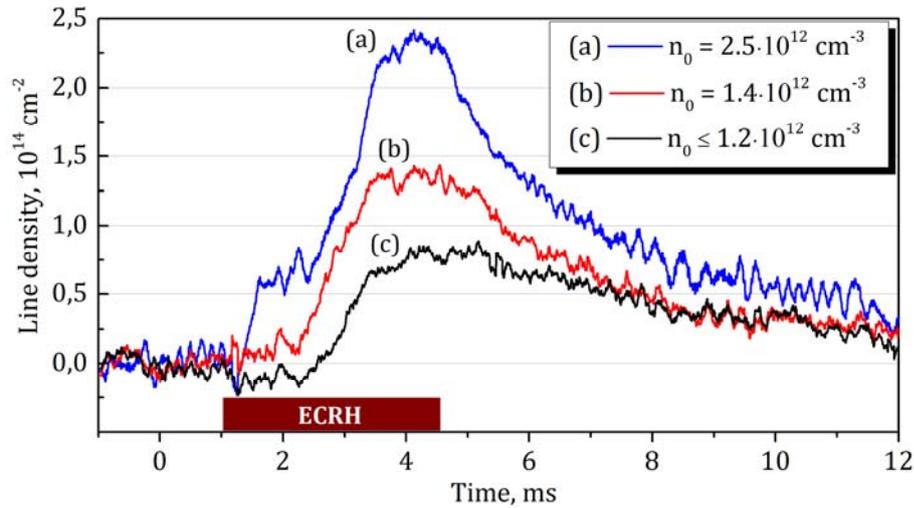

**Figure 8.** Line-averaged density of seed plasma with different initial densities of neutral gas. Non-physical negative electron density observed before the breakdown is within the experimental accuracy of $\pm 0.2 \times 10^{14}$ cm$^{-2}$.

## 4. Main discharge stage with NBI heating

The ultimate goal for the ECR startup is to build up a plasma with sufficient density, which enables accumulation of fast ions during NBI. In this section we test how suitable is the ECR-generated plasma target in GDT conditions using the 5 MW NBI heating system.

The main tool used for measurement of accumulate fast ion energy is the diamagnetic loop encircling the fast ion reflection point (figure 1). Typical diamagnetic signal of fast ions with ECR-initiated discharge is shown in figure 9 (bottom panel); corresponding line density is shown in figure 10. The ECR-initiated discharge is compared against a gun-initiated discharge with similar experimental conditions. As follows from the captured NBI power in plasma (figure 9, top panel) and diamagnetic signal, the first 1-2 ms are very different from a gun discharge. With the start of NBI, the captured power fraction starts with a small value < 10% and gradually arrives at a stationary value of ~40%, whereas for a gun-initiated discharge plasma thickness usually starts at ~50% and falls off further, which is also reflected by the line density signal. Note that the density drop (figure 10) is explained by shock "inflation" of gun plasma, which is known to be rather thin compared to NBI power deposition profile. On the other hand, with ECR-generated plasma target, only a steady growth of line density is observed.

With a dedicated series of discharges we tried to estimate the threshold for accumulation of fast ions. Starting from the optimum regime of the ECR plasma breakdown, the gyrotron pulse was gradually curtailed, effectively limiting the density of seed plasma. The gas feed scenario was maintained to provide identical conditions for plasma support during NBI. The decrease in seed plasma line density at first leads only to prolongation of transient stage (figure 11). However, after a borderline value of $\sim 0.5 \cdot 10^{14}$ cm$^{-2}$ plasma can decay entirely, even before NBI pulse termination. The fraction of NBI power captured in plasma in this borderline case is ~5% or ~250 kW. The values slightly above the threshold in principle allow a reliable discharge startup, but with the NBI pulse available for the GDT experiment such delayed startup is not acceptable. Raising the line density value to at least $1.5 \cdot 10^{14}$ cm$^{-2}$ leads to reasonably short transient stage, which is similar to the standard gun discharge (curve (a) in figure 9).

The general plasma parameters with ECR breakdown are mostly similar to a gun-initiated discharge. On-axis electron temperature is typically within 180-200 eV, which is somewhat lower than usual 200-250 eV (without additional ECRH). In compared discharges (figure 9) the plasma energy content is ~10% higher with ECR breakdown, but the actual NBI energy absorbed by plasma is ~15% lower. A slight on-axis temperature difference and equal energy loss rate (figure 9, center panel), indicate a difference in plasma density and temperature profiles. The fact about effectively improved energy confinement is explained by excessive energy loss during the transient stage in a gun-initiated discharge.

As it was expected, the vacuum conditions in the end tanks are essentially different with ECR breakdown. During the gun operation, only a fraction neutral gas supplied to the arc channel is converted to plasma, while the rest of it is being spread over the whole tank volume. With active plasma gun the neutral $D_2$ density in the gun tank reaches $\sim 7 \cdot 10^{13}$ cm$^{-3}$ before the NBI pulse. In contrast, with ECR startup the neutral gas density stays below $10^{11}$ cm$^{-3}$ for the whole plasma discharge.

Despite these improvements, we did not observe any detectable effect on longitudinal heat conduction with ECR startup. However, the effect manifested itself in increased requirements for vortex stabilization. As shown in figure 12, to achieve plasma stabilization the limiters are biased with twice the voltage applied in a gun regime. Even with this, the limiter current ends up noticeably lower. This signifies a substantial decrease in plasma conductivity between the end-plate and the limiter, which is a consequence of improved vacuum conditions in the end-tanks. This and other effects of neutral pressure in the region of magnetic field expansion are reserved for a dedicated study.

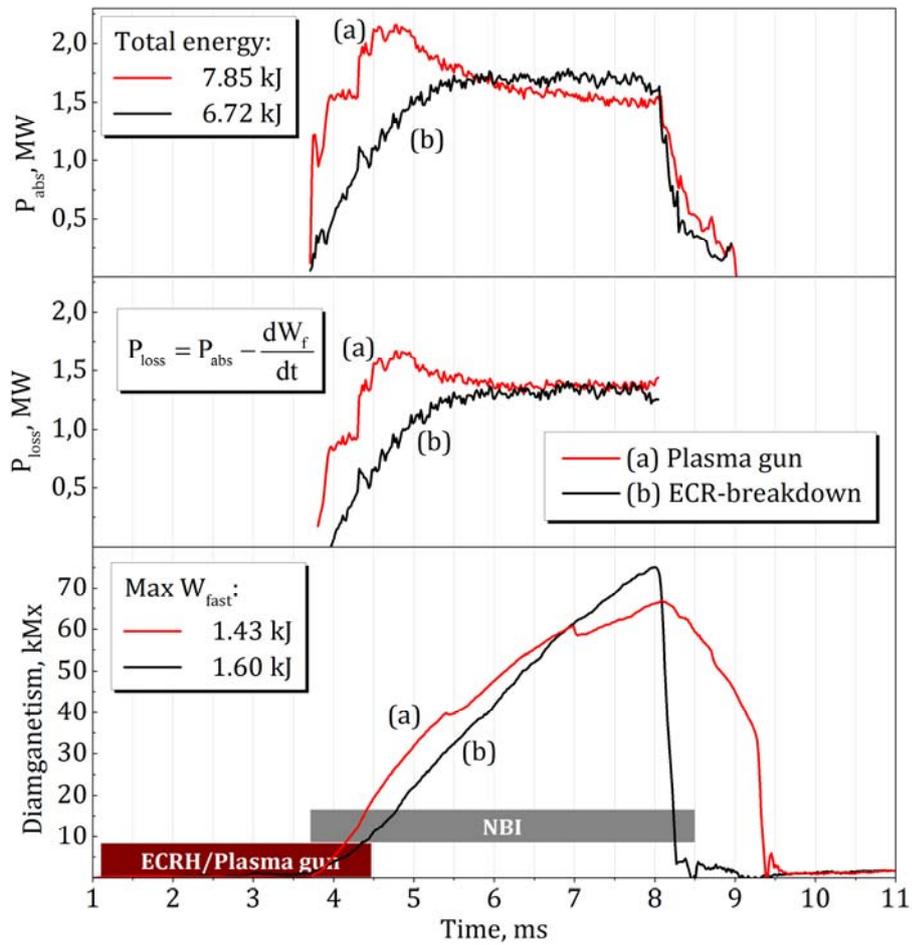

**Figure 9.** Diamagnetic signal of fast ions (bottom), captured in plasma NBI power (top) and calculated power loss (center) for gun- and ECR-initiated discharges (cases a) and b), respectively).

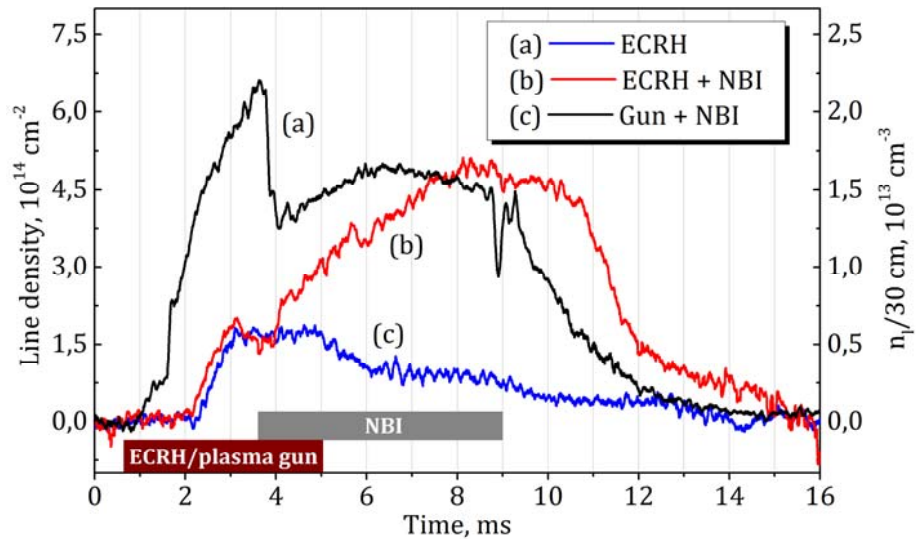

**Figure 10.** Line-averaged density of ECR-generated plasma with and without NBI; line plasma density in a gun-initiated discharge.

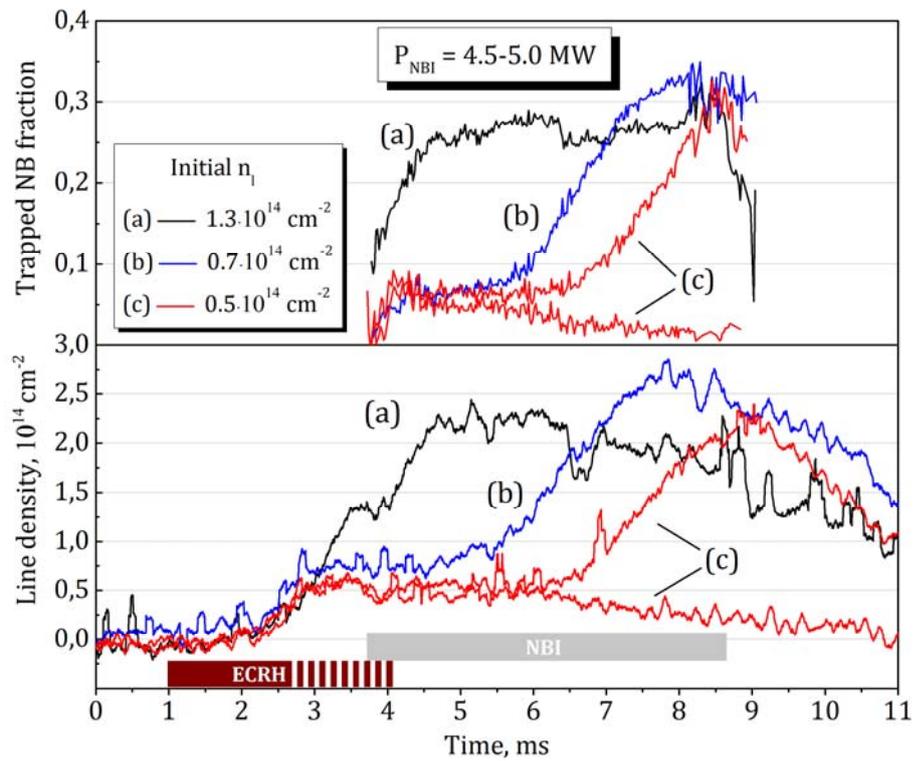

**Figure 11.** Captured NBI power (top) and line-averaged plasma density (bottom) in discharges with variable microwave pulse duration. (a), (b) – reliable discharge startup. (c) – high probability of plasma decay

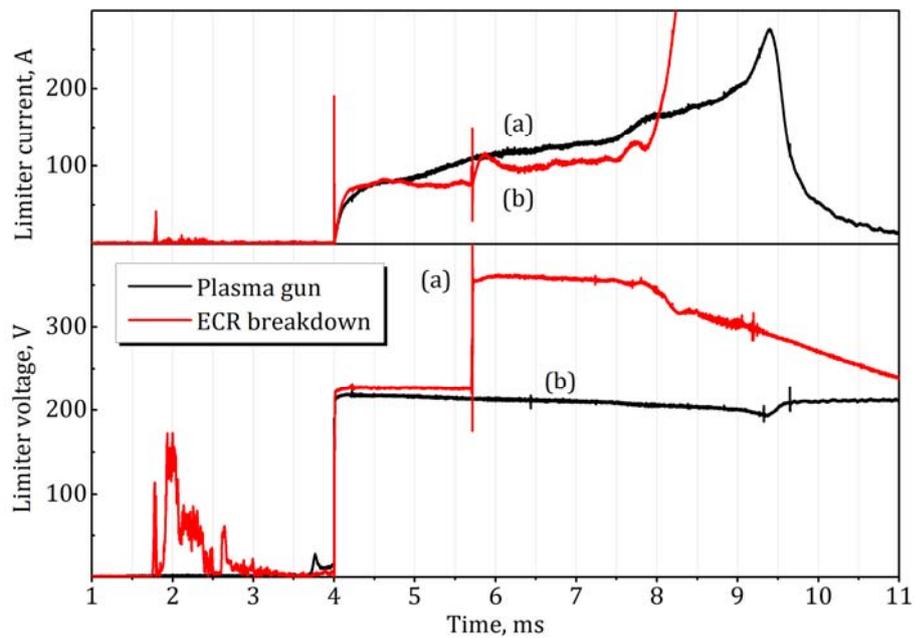

**Figure 12.** Limiter voltage (bottom) and current (top) with gun- and ECR-initiated discharges.

## 5. Discussion of the results. Theory of ECR start-up in GDT

Before proceeding to analysis, let us summarize the essential facts about ECR seed plasma generation. Irradiation of fundamental harmonic ECR surface by a microwave beam with power > 50 kW leads to breakdown of neutral gas and generation of plasma opaque to microwaves. Experiment with variable magnetic field shows that density buildup is not sensitive to intersection of ECR surface by microwave beam. For values of incident power above ~150 kW, the line density evolution is identical and is determined by the initial gas pressure. At lower values, the dependence on microwave power is observed. Typical microwave pulse duration needed to reach the saturated line density is ~2 ms.

*5.1. Basic scenario*

To start theoretical interpretation of the experimental findings let us consider a simple ionization balance equation,

$$\frac{dn_e}{dt} = n_a n_e \langle \upsilon \sigma_{ion} \rangle, \qquad (1)$$

where $n_a$ is the density of neutral atoms, i.e. the doubled density of the initial gas, $n_a = 2n_0$; $n_e$ is the electron density; $\langle \upsilon \sigma_{ion} \rangle$ is the ionization constant averaged over the electron distribution function, and all electron losses are neglected. This equation may be solved for unknown electron density $n_e$ required to maintain the observed plasma density growth rate $dn_e/dt$ at the plasma build-up phase. The result is shown in figure 13. Let us take, for instance, the growth rate and neutral deuterium density from the discharge shown in figure 6:

$$\frac{dn_e}{dt} \approx \frac{3\times10^{12}\,\text{cm}^{-3}}{10^{-3}\,\text{s}}, \quad n_a \approx 3\times10^{12}\,\text{cm}^{-3}.$$

The ionization constant is calculated for two different models of the electron distribution function—Maxwellian distribution function with the temperature $T_e = E^*$, and the quasi-linear plateau formed in the region $E < E^*$ under strong ECR heating as described below. The latter model is more adequate for electrons with mean energy above 100 eV and corresponds to low collisionality.

For measured line plasma density $1.8\times10^{14}\,\text{cm}^{-2}$ one may assume a uniform plasma with diameter of 60 cm, which gives

$$n_e \approx 3\times10^{12}\,\text{cm}^{-3}.$$

The reason for this choice is the TS data, which puts an upper limit on plasma density, and the fact that 60 cm is the diameter of diamagnetic loop; survival of plasma beyond the loop is highly unlikely. The electron density level is shown as a horizontal line in figure 13. The solution suggests that ionization balance is possible in highly collisional plasma with $T_e < 10\,\text{eV}$, see point (a) in figure 13. However, this simple assumption is inconsistent with limitations extracted from the diamagnetic loop measurements, which are projected as a gray strip in figure 13.

Indeed, assuming that only the electrons contribute to the diamagnetic flux during the plasma start-up, the total transverse pressure (to the confining magnetic field) at the midplane evaluates to:

$$E_\perp n_e \approx B\,\Delta\Phi/S_0,$$

where $E_\perp$ is the mean transverse energy of electrons, $\Delta\Phi \approx 1\,\text{kMx}$ and $B = 0.35\,\text{T}$ are the diamagnetic flux and magnetic field at the midplane, respectively, and $S_0$ is the plasma cross section. For collisional electrons with isotropic distribution function $E = E_\perp$. For adiabatically confined electrons, which are reflected back at ECR layer (as it is suggested by the diamagnetic signals) one can estimate the mean energy as $E = R_{ECR} E_\perp$ where $R_{ECR} = B_{ECR}/B_{center} = 1.95\,\text{T}/0.35\,\text{T}$ is the mirror-ratio corresponding to ECR. These two limiting cases form the lower and upper boundaries of the gray zone in figure 13. Assuming that the diamagnetic signal comes from the bulk plasma with the temperature $T_e \approx 20\,\text{eV}$, see point (b) in figure 13, we immediately find that ionization rate cannot be supported by the same plasma.

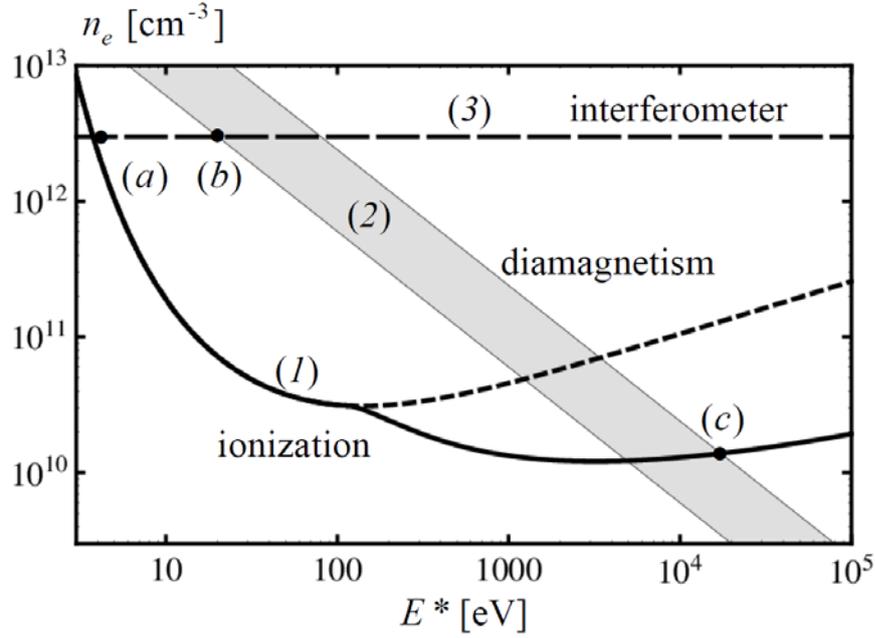

**Figure 13.** Limitations on the density $n_e$ and characteristic energy $E^*$ of electrons set by experimental data. (*1*) Solution of Eq. (1) for $n_e$ with density growth rate $dn_e/dt = 3\times10^{15}\,\text{cm}^{-3}\text{s}^{-1}$ and $n_a \approx 3\times10^{12}\,\text{cm}^{-3}$; the fork at $E^* \approx 100$ eV corresponds to a transition from Maxwellian distribution function (upper branch) to the collisionless quasi-linear plateau (lower branch). (*2*) The parameter region consistent with the diamagnetic loop data. (*3*) The density level $n_e = 3\times10^{12}\,\text{cm}^{-3}$ recovered from interferometer data.

The natural solution that matches both the interferometer and diamagnetic data is based on the assumption that ionization occurs on much more energetic, but less populated electron fraction (point (c) in figure 13). The hot collisionless electrons with strongly anisotropic distribution function, mean energy $\sim 10$ keV, and density $\sim 10^{10}\,\text{cm}^{-3}$ in the main body of the trap, bounce between the opposite ECR layers and ionize the background gas on their passage through the trap. The secondary electrons are accumulated in the main body of the trap and essentially do not reach the ECR zone. Therefore their temperature remains as low as few eVs, but the density can reach the level measured by the interferometer because of high ionization rate by hot electrons. Due to the low temperature, the secondary electrons do not contribute to the diamagnetic flux, which is fully determined by the hot anisotropic electrons. In figure 13, the secondary electrons occupy the region between points (a) and (b). Below we develop a quantitative model to support this scenario of plasma buildup.

*5.2. Formation of a hot electron population*

The most suitable to GDT conditions theory of ECR breakdown of rarefied gas in a mirror trap was proposed in [31], and later was verified experimentally in [8]. The theory is based on the following assumptions:
(a) The influence of the collisional scattering of electrons by particles is negligibly small.
(b) The electrons are heated stochastically while bouncing along a magnetic field line between magnetic mirrors and passing the localized ECR zone at one trap end. The average change in the kinetic energy of electrons in their passing through the ECR zone is large compared to the energy of electrons leaving the trap.
(c) The electrons gain mostly "transverse" kinetic energy stored in cyclotron gyration, that exceeds by many orders of magnitude the ionization energy $E_{\text{ion}}$. The initial energy of the newborn electrons is of the order of the ionization energy. The kinetic energy of the longitudinal motion in the ECR region slightly varies due to interaction with waves, so relation $E_{ion} \lesssim E_{\parallel} \ll E_{\perp}$ holds during the discharge.

(d) The kinetic energy of electrons during the ECR heating has an upper limit, $E_{\perp} < E^*$, above which the acceleration stops.

All these conditions are valid in the GDT plasma.

The main distinction from the cited works is related to the mechanism responsible for the upper limit of the electron energy. Instead of so-called super-adiabatic limit typical of systems with longitudinal wave propagation, in our case the maximal electron energy is limited by the shift of the EC resonance towards the turning point of the accelerating electrons in rare plasma. Namely, the maximal energy is

$$E^* \approx \sqrt{\frac{2mc^2 E_{ion}}{1-n_{\parallel}^2}} \quad , \qquad (2)$$

where $E_{ion} \approx 15$ eV is the ionization energy for deuterium, $mc^2 = 512$ keV is the electron rest energy, and $n_{\parallel} = k_{\parallel} c / \omega$ is the longitudinal index of refraction for electromagnetic waves. This formula is derived in appendix A1. In GDT, radiation with frequency 54.5 GHz is launched at $\theta = 36°$ to the magnetic field; considering vacuum value $n_{\parallel} = \cos \theta$ one finds $E^* = 6.5$ keV. This value is consistent with our basic scenario.

It should be noted, however, that Eq. (2) is obtained under assumption that longitudinal and perpendicular velocities of the resonant electron change along the quasi-linear diffusion lines, see Eqs. (A4) and (A5) in appendix. For a minor fraction of hot electrons, this condition may be violated allowing acceleration up to much higher energies. In this case the maximal energy is defined by a finite width of the microwave beam along the magnetic field,

$$E^{**} \approx 2mc^2 \delta B_{wave} / B ,$$

where $\delta B_{wave} / B$ defines the relative variation of the magnetic field between the "cold" resonance point, $\omega_B = \omega$, and the edge of the wave beam. For the GDT conditions $\delta B_{wave} / B \sim 0.1$, which results in the upper limit $E^{**} \approx 100$ keV. The same limitation holds in dense enough plasma with $n_{\parallel} > 1$, where Eq. (2) is no longer valid [32]. The critical plasma density corresponding to a transition to the dense plasma regime of electron acceleration may be estimated from

$$n_{\parallel}^2 \approx \left(1 + \frac{\omega_p^2}{n_{\parallel} \omega^2 \langle \beta_{\parallel} \rangle_{ECR}}\right) \cos \theta \approx 1 ,$$

where $\omega_p$ and $\langle \beta_{\parallel} \rangle_{ECR} = \langle v_{\parallel} / c \rangle_{ECR} \sim \sqrt{E_{ion} / mc^2}$ are the electron Langmuir frequency and the mean longitudinal velocity of hot electrons in the ECR zone. For the GDT conditions this gives density of hot electrons $n_{h,ECR} \sim 2 \times 10^{11} \text{cm}^{-3}$ in the ECR zone. Density pinching due to close reflections of the anisotropic electrons inside the ECR zone is estimated as $\sqrt{E^*/E_{ion}} \sim 20$. Therefore, the energy limit (2) seems to be marginally compatible with our model even taking into account the pinching of the hot electrons.

Under conditions (a)-(e), formation of the distribution function of hot electrons in phase space is governed by the Fokker–Planck kinetic equation [31-36]. This process occurs in a strongly collisionless regime in which the interaction with microwaves defines both the acceleration of newborn electrons and their escape into a loss-cone. Leaving details for the Appendix, we present the main results of a kinetic approach. Typical time required for acceleration of a newborn electron up to the maximal energy $E^*$ is about 1 μs, which is much shorter than all times related to the plasma density evolution and microwave pulse length. Under these circumstances, the distribution function forms the "quasi-linear plateau" in resonant region of the momentum space. For simplicity we use the one dimensional distribution function over the total kinetic energy, $F(E)$, for which the following approximate solution of the kinetic equation may be found (appendix A2):

$$F_{\text{pl}}(E) = \frac{N(t)}{2\sqrt{EE^*}} \begin{cases} 1, & 0 < E < E^* \\ 0, & E > E^* \end{cases}. \quad (3)$$

where $N(t)$ is the total number of particles in a flux tube, and index stands for "plateau". Note this distribution function is applicable at any point along the bounce trajectory since $E$ is an invariant of the adiabatic motion outside the ECR region. Therefore, this distribution allows determining the quantities averaged over the whole plasma volume (not only inside the heating zone). This way we obtain a volume averaged ionization rate (appendix A3):

$$\langle \upsilon \sigma_{\text{ion}} \rangle_{\text{pl}} = N^{-1} \int_{E_{\text{ion}}}^{E^*} \sigma_{\text{ion}} F_{\text{pl}} \sqrt{2E/m}\, dE \approx \frac{3.1 \times 10^{-7}\, \text{cm}^3/\text{s}}{\sqrt{E^*[\text{eV}]}} \left(0.9 + \ln^2 \frac{E^*}{E_{\text{ion}}}\right). \quad (4)$$

The last relation corresponds to the particular case of deuterium. One can find that in this case the ionization rate weakly varies around $\langle \upsilon \sigma_{\text{ion}} \rangle_{\text{pl}} \approx 6 \times 10^{-8}\, \text{cm}^3/\text{s}$ in a wide range of limiting energies $E^*$ from 1 keV to 100 keV.

Another important quantity is the loss rate of the hot electrons. In a rarefied plasma these losses may be attributed to particle flux induced by waves into a loss-cone [37] and may be found as follows. First, the quasi-linear plateau is assumed in the whole two-dimensional momentum space including the loss-cone region; then, the corresponding particle flux into the loss-cone is calculated. The result for a long trap in the approximation of high anisotropy of the electron distribution function in the resonance region ($E \approx E_\perp \gg E_\parallel$) may be expressed as (appendix A4)

$$\frac{1}{\tau_h} \equiv \frac{1}{N}\left(\frac{dN}{dt}\right)_{\text{loss}} \approx \frac{\sqrt{1 - B_{\text{center}}/B_{\text{ECR}}}}{B_{\text{plug}}/B_{\text{ECR}} - 1} \frac{\langle E_\parallel \rangle_{\text{ECR}}}{L\sqrt{2mE^*}}. \quad (5)$$

Here $B_{\text{plug}}$, $B_{\text{ECR}}$ and $B_{\text{center}}$ are the magnetic fields at the mirror plugs, the ECR zone and the trap center, correspondingly; $L$ is the mirror-to-mirror length; $\langle E_\parallel \rangle_{\text{ECR}} \approx E_{\text{ion}}$ is the mean longitudinal energy at the ECR surface. For the GDT conditions, the loss rate of hot electrons induced by strong microwaves is about $\tau_h \approx 0.5$ ms. Accidentally, this value is close to the decay time of the hot electrons after ECRH switch-off, estimated from the hard X-ray signals (figure 6). Note that hot electron losses, both during and after the ECRH stage, may also be related to development of stimulated electron cyclotron instabilities [38].

Within the proposed quasi-linear model there is no principal difference between generation mechanisms of hot electrons in different magnetic configurations, provided that the ECR is present along a field line, in particular, between cases (b) and (c) shown in figure 3. In both cases, confined hot electrons bounce between the ECR surfaces and do not feel the outside region. However, for the peripheral ECR corresponding to case (c), part of the field lines ends at the limiter, which results in a reduced effective mirror ratio, i.e. if the whole ECR zone is in the shadow of the limiter, one should use $B_{\text{limiter}} \sim 1.25 B_{\text{ECR}}$ instead of $B_{\text{plug}} \sim 5 B_{\text{ECR}}$ in Eq. (5). This increases the flux of lost electrons escaping from the trap along the field lines as compared to the on-axis ECR. Moreover, born with the peripheral ECR, the electrons bombard the limiter and likely cause the bremsstrahlung X-rays registered by a closely standing detector, while with the on-axis ECR they go far away to the expander tank.

Regardless of the details of hot electron losses, the ECR breakdown condition

$$n_a \langle \upsilon \sigma_{\text{ion}} \rangle_{\text{pl}} > 1/\tau_h,$$

is easily fulfilled in the experiment: $n_a > 3 \times 10^{10}\, \text{cm}^{-3}$. This also might explain the experimentally observed weak dependence of discharge on incident microwave power. However, our case is slightly more complicated than the common avalanche mechanism—the secondary electrons are born outside the ECR zone, so they do not contribute directly to the avalanche. A model of such situation is developed below.

*5.3. Density saturation mechanism*

To proceed further we need to define the losses of the bulk electrons. Indeed, this is the most uncertain part of our theory. We must start from the well established experimental fact that during the ECR breakdown the density of seed plasma goes to the saturated value that is constant until the microwave power is switched off. The saturated level is roughly proportional to the initial neutral gas density; and the growth rate seems to be also proportional to the gas density, however experimental accuracy does not allow verifying it exactly.

We failed to explain the density saturation with any electron loss mechanism known from first principles, so we propose the following phenomenological model. Let us assume that ECRH essentially reduces the bulk plasma losses such that the number of charged particles and neutrals in the whole trap is approximately conserved. Any other than ionization losses or sources of a neutral gas are neglected as well. Then, the ionization stops when all neutrals are consumed; and the flattop plasma density is defined only by initial number of neutrals. This simple assumption allows reproducing most of the experimental observations quite well. As a first example, figure 14 shows the approximation of the interferometer data with a solution of the simplest balance equation (1) with $n_e + n_a = \text{const}$ and $\langle \upsilon \sigma_{\text{ion}} \rangle = \text{const}$. More sophisticated realization of the proposed idea is explained in the next section.

Note that the physical reason for the ECR plugging is still not clear. The density of the splashing hot electrons seems to be not enough for ambipolar plugging of the main plasma. On the other hand, the pondermotive force acting on the bulk electrons may, in principle, plug them as described in [39]. Indeed, the pondermotive potential $\varphi \approx e^2 E_-^2 (2m\omega\Delta\omega)^{-1}$ becomes of the order of $E_{\text{ion}}$ for the cyclotron frequency detuning $\Delta\omega = \omega - \omega_B$ far before the ECR zone for the hot electrons. However, in our conditions the pondermotive ECR plug acts only at one end of the trap.

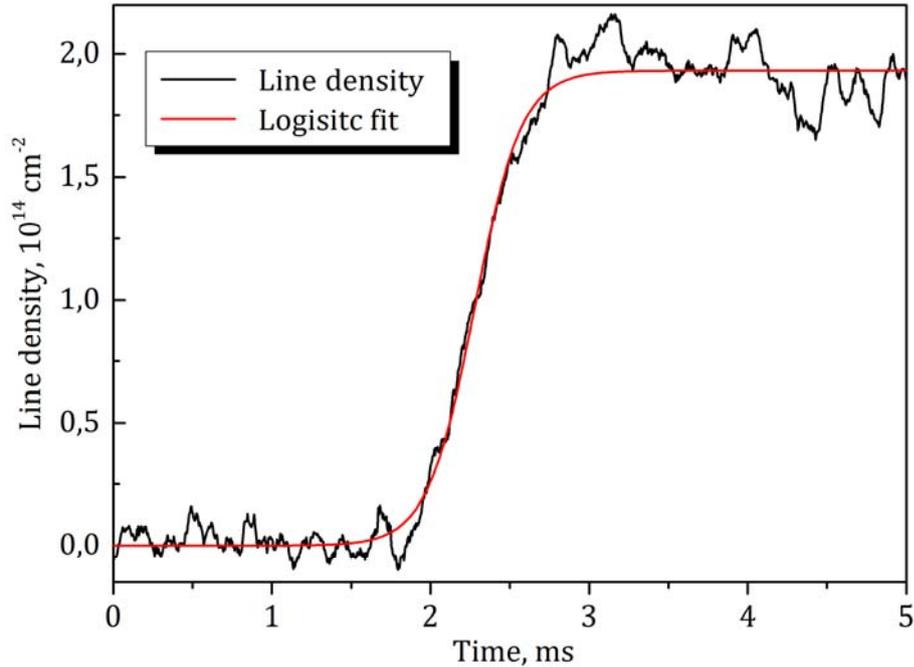

**Figure 14.** Approximation of interferometer data shown in figure 6 with a logistic function.

*5.4. Particle and energy balance*

Here we present rather accurate modeling based on the idea of a discharge with two distinct electron components. We solve the following set of balance equations for the bulk density of thermal electrons $n_e$, the density of hot electrons calculated at the trap center $n_h \approx N(t)/V_{\text{plasma}}$, the atomic density of neutrals $n_a$, and the temperature of bulk electrons $T_e$:

$$\begin{cases} \dfrac{dn_e}{dt} = \kappa n_h n_a \langle \upsilon \sigma_{\text{ion}} \rangle_{\text{pl}} - \dfrac{n_e}{\tau_e} \\ \dfrac{dn_h}{dt} = g(t)\dfrac{n_e}{\tau_e} - \dfrac{n_h}{\tau_h} \\ \dfrac{dn_a}{dt} = -\kappa n_h n_a \langle \upsilon \sigma_{\text{ion}} \rangle_{\text{pl}} \\ \dfrac{3}{2}\dfrac{d}{dt}(n_e T_e) = n_h \langle \nu_{eh} E \rangle_{\text{pl}} + \kappa n_h n_a \langle \upsilon \sigma_{\text{ion}} \rangle_{\text{pl}} E_{\text{sec}} - \dfrac{5 n_e T_e}{\tau_e} \end{cases} \quad (6)$$

It is implied in this model that all bulk electrons lost along the magnetic field lines during the microwave heating, finally reach the ECR region and gain energy on a fast time scale. This is modeled by the same term $n_e/\tau_e$ acting as a sink for the cold electrons and a source for the hot electrons. Here function $g(t) = \exp[-(t/\tau_{\text{gyr}})^{32}]$ models the fast switch-off of gyrotron at time $t = \tau_{\text{gyr}}$; when $g = 0$ there is no transfer from cold electrons to hot electrons. Particular value of the bulk plasma confinement time $\tau_e$ during the ECR stage is defined by fit to the experimental data—for a known initial gas density and plasma growth rate we determine the required density of hot electrons $n_h^*$ and the flattop density of the bulk plasma $n_e^* = n_a(0)$; these two quantities are compatible if $n_e^*/\tau_e \approx n_h^*/\tau_h$. When the gyrotron is off, the cold electrons are confined in the gas dynamic regime [40]. Combining these two confinement regimes we propose the following model equation for the cold electron loss rate:

$$\frac{1}{\tau_e} = g(t)\frac{n_h^*}{n_e^*}\frac{1}{\tau_h} + (1-g(t))\frac{2}{RL}\sqrt{\frac{T_e}{M}}$$

with $M$ being the ion mass and $R = B_{\text{plug}}/B_{\text{center}}$ is the trap mirror ratio. On the other hand, the ionization term acts as a source for the bulk electrons only, so the direct avalanche of hot electrons is avoided, and as a sink for the neutrals; here $\kappa \approx R_{\text{ECR}}$ is a constant geometric factor. The last equation in (6) is the energy balance for the bulk electrons. Terms in the r.h.s. are, respectively, the power transfer due to collisional friction of hot electrons, the gain due to newly born secondary electrons with effective energy $E_{\text{sec}} \approx \frac{3}{2}E_{\text{ion}}$ (see Appendix A5), and the longitudinal losses in the gas-dynamic regime. In calculations we also take into account power losses due to electron impact excitation of line radiation of the bulk ions and neutrals (not shown here for simplicity).

The results of typical discharge modeling are presented in figure 15. Initial parameters are close to the conditions in figures 4 and 6. One can see that the model, actually featuring only one free parameter $\tau_e$, may reproduce the experimental data fairly well—compare the bulk and hot densities, $n_e$ and $n_h$, correspondingly, with the interferometer and X-ray signals in figure 7. One can see that plasma density saturation is accompanied by the depletion of the neutral gas; this stage can be maintained as long as microwaves plug the bulk plasma losses. The bulk electron temperature varies from 3 eV to 8 eV (figure 15, right), thus being below the 20 eV level required for bulk plasma to be visible in the diamagnetic measurements.

Although we do not calculate a full power balance for the hot electrons, we can find the main energy losses as

$$P_h \approx \left( \langle E \rangle_{\text{pl}} dn_h/dt + \langle E/\tau_b \rangle_{\text{pl}} + \langle \nu_{eh} E \rangle_{\text{pl}} + \kappa n_a \langle \upsilon \sigma_{\text{ion}} \rangle_{\text{pl}} (E_{\text{ion}} + E_{\text{sec}}) \right) n_h V_{\text{plasma}}. \quad (7)$$

The terms in the r.h.s. are, respectively, the power for creating the hot electrons with the mean energy $\langle E \rangle_{\text{pl}} = \frac{1}{3} E^*$ corresponding to the distribution function (3), the energy losses of hot electrons associated to the flux into the loss cone, the losses due to collisional friction of hot electrons with the bulk plasma, the ionization losses and the losses that come to the secondary electrons. Details on calculating different loss channels may be found in Appendix A5. The result of calculations (see figure 15, right) is that the total discharge power is about 30 kW, thus being far less then the injected gyrotron power of about 400 kW. This estimate suggests the critical level below which the discharge evolution depends on the deposited power; however the experimentally observed level is five times higher (150 kW). Possible reasons of such discrepancy are mentioned in the next section.

To demonstrate the flexibility of the model, in figure 16 we reproduce the experiments with the gas pressure scan reported in figure 8.

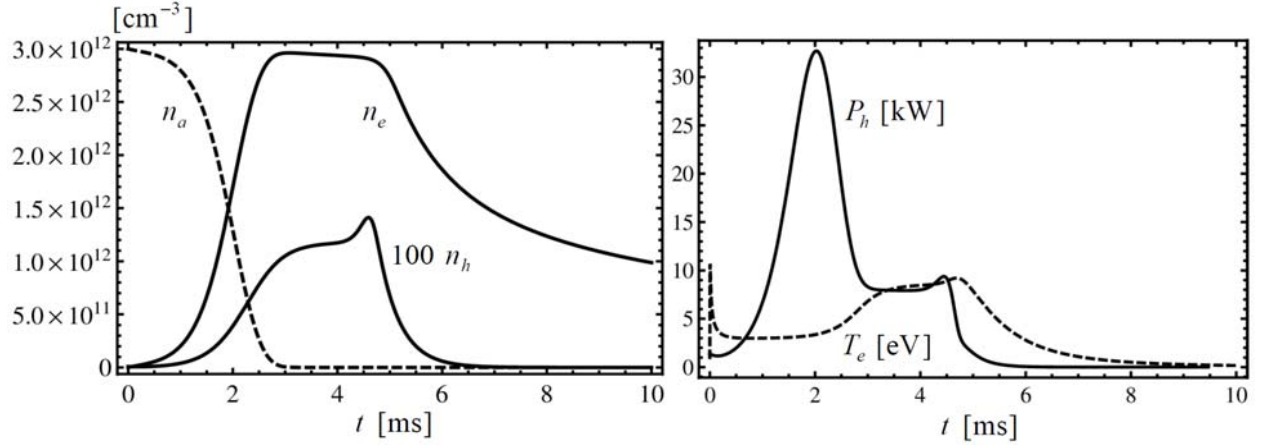

**Figure 15.** The discharge modeling based on Eqs. (6): evolution of densities (left), electron temperature and total power lost (right). Parameters of calculations are $n_a(0) = 3 \times 10^{12}\,\text{cm}^{-3}$, $n_h^* = 1.2 \times 10^{10}\,\text{cm}^{-3}$, $\tau_{gyr} = 5\,\text{ms}$, $E^* = 6.5\,\text{keV}$, $\kappa = 5$, $L = 7\,\text{m}$, and plasma diameter is 30 cm.

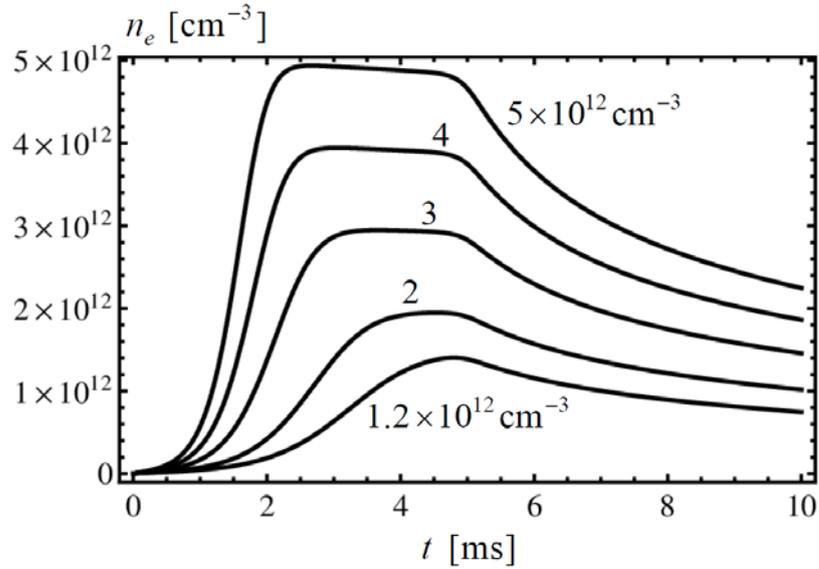

**Figure 16.** Modeling of the seed plasma density evolution for different initial gas densities from $n_a(0) = 1.2 \times 10^{12}\,\text{cm}^{-3}$ to $5 \times 10^{12}\,\text{cm}^{-3}$. Other parameters are the same as in figure 15.

### 5.5. Microwave power absorption

As suggested by stray radiation monitors, essential part of the injected microwave power is absorbed during the entire stage of the breakdown except for the initial phase. This is supported by simple estimate of the resonant power absorption from the radiative transfer equation along the ray $l$

$$dI/dl = -2\,\text{Im}\,k\,I,$$

where $I$ is the wave beam intensity, $2\,\text{Im}\,k$ is the absorption coefficient. For a rough estimate, the main contribution to absorption coefficient may be calculated within cold plasma approximation for waves propagating quasi-longitudinally with respect to the external magnetic field (see, e.g., Ch. 3 § 11 in Ref [41]; § 3.4 problem 3.8 1 in Ref.[42])

$$2\,\mathrm{Im}\,k = \frac{\nu}{c}\frac{\omega_p^2}{(\omega-\omega_B)^2 + \nu_{\mathrm{eff}}^2},$$

where $\omega$, $\omega_p$ and $\omega_B$ are the wave, electron Langmuir and cyclotron frequencies, correspondingly, and $\nu_{\mathrm{eff}}$ is the effective absorption rate. Let us assume now that $\omega - \omega_B \approx \omega l \cos\theta / L_B$ is linearly varying in the vicinity of the cyclotron resonance; here $L_B = B/|\nabla B| \approx 100$ cm is the scale of magnetic field variation along the trap axis, and $\theta = 36°$ is the angle of a straight ray $l$ to the axis. Then, the total absorption is independent of particular value of $\nu_{\mathrm{eff}}$ as long as $\nu_{\mathrm{eff}} \ll \omega$:

$$I_{\mathrm{out}}/I_{\mathrm{in}} = \exp(-\eta)\,, \quad \eta = \int_{in}^{out} 2\,\mathrm{Im}\,k\, dl \approx \pi \omega_p^2 \frac{L_B}{c\omega\cos\theta}. \qquad (8)$$

The same result may be obtained for an exact solution of a model wave equation, see Ch. 5 § 27 [41] or § 6.2.3 [42]. Plasma becomes opaque when $\eta > 1$. For GDT this corresponds to the extremely low electron density limit $> 10^{10}\,\mathrm{cm}^{-3}$.

Therefore, full absorption of the left-hand polarized wave (extraordinary mode) is possible for the entire course of the discharge except for the very beginning. However, our modeling predicts the net power consumed by the ECR discharge of the order of 10% of the injected microwave power (figure 15, right). This doesn`t seem unnatural for a high-power discharge in a rarefied gas since the formation of the collisionless quasi-linear plateau in the distribution function of hot electrons results in degradation of cyclotron damping [33, 37] and in eventual expansion of energy deposition profile [43]. Note that low efficiency of conversion of RF power to hot electrons was previously observed in TMX-U experiment [6]. In a somewhat similar to GDT regime, ECRH was used both for initial breakdown of gas and for generation of hot electrons. In contradiction to the direct measurements of the absorbed power and numerical simulations, which predicted near full absorption, the typical energy growth rate of hot electrons could only account for a mere 15% of incident microwave power.

In case of GDT, we suppose that the rest of the microwave power is absorbed by chamber walls and low density residual plasma that forms at the periphery. Absorption in the residual plasma seems to be the most reasonable explanation of the observed power balance of the discharge. The same situation was faced previously in occasional unsuccessful shots in experiments with ECR heating of the main plasma, when no increase of electron temperature was detected simultaneously with the stray radiation decrease. The periphery plasma is poorly confined and remains cold, but can still absorb the microwaves provided its density is at least $10^{10}\,\mathrm{cm}^{-3}$. Another issue that is possible at such density levels is the reflection of gyrotron radiation from a tenuous plasma in the launching port at the parasitic ECR resonance as described in section 3.1. It must be admitted, that the full electrodynamics study of these processes is beyond the scope of present paper.

## 6. Conclusion

A new discharge startup scenario has been developed for the GDT magnetic mirror experiment. The initial plasma target is generated from neutral gas with a help of a high-power microwave beam available from the ECR plasma heating system. The approach was extensively tested in the conditions of a large-scale axisymmetric magnetic mirror machine with neutral beam injection. In the experiments we scanned the ECR layer with the microwave beam, studied the dependence on the incident microwave power, polarization and initial neutral gas pressure. It was found that the seed plasma density saturates at a certain level depending on the amount of neutral gas initially admitted to the vessel, in case the power is sufficiently high. Below a certain value, the incident microwave power becomes a limitation to plasma density. Positioning of the microwave beam relative to the ECR layer does not affect the plasma buildup, as long as the beam intersects the first harmonic ECR surface.

Large collection of experimental data allowed us to develop a theoretical interpretation of not only the initial stage of gas breakdown, but also the subsequent accumulation of seed plasma. We propose a two-component discharge concept, in which ECR heating produces a minor fraction of hot electrons ($E \sim 10\,\text{keV}$, $n_h \sim 10^{10}\,\text{cm}^{-3}$) which are fully responsible for gas ionization and plasma pressure at the plasma build-up stage. The secondary electrons, which are produced during ionization, form a cold and dense seed plasma ($T_e \sim 5\,\text{eV}$, $n_e \sim 10^{12}\,\text{cm}^{-3}$) characterized by improved confinement during ECRH. Physical mechanism of such confinement is still a challenging question for further research.

A threshold value of line-averaged density $\sim 0.5 \cdot 10^{14}\,\text{cm}^{-2}$, which permits reliable startup of a discharge in GDT, was determined experimentally. It must be noted, however, that a low line density is only an indication of general "weakness" of seed plasma and the premature loss of fast ions may only be explained by the charge exchange with residual neutral gas. The transition from ECR-generated plasma to purely NBI-supported discharge was shown to be sufficiently short and even more energy saving than the conventional startup with arc plasma generator. After the transition, the discharge becomes essentially similar to a one initiated by the plasma gun. It can be concluded that a sufficiently dense plasma target in GDT may be generated using a 50 kW microwave beam directed at the fundamental EC resonance with the minimum duration of ECRH pulse of $\sim 3$ ms.

Finally, our experience with ECR-initiated discharges leads us to conclusion that such startup scenario will be easily achieved in a next-generation magnetic mirror machine based on similar physics. It should be stressed that despite the lack of external stabilization against the interchange modes during the ECRH pulse, the transverse energy transport does not seem to be extraordinary high, leaving a significant power budget to spare and allowing a reliable plasma startup in each discharge.

**Acknowledgements**

The authors thank Prof. S. V. Golubev for his constant encouragement and support, Prof. M. D. Tokman, whose works highlighted the essential role of quasi-linear effects in ionization and confinement of fast electrons in mirror trap, and the GDT team which carried out the experimental part the work. The work is supported by the Russian Science Foundation (project No 14-12-01007).

**Appendix. Collisionless dynamics of electrons during cyclotron heating in a magnetic mirror**

*A1. Maximum energy*

Let us consider adiabatic movement of an electron bouncing along a magnetic field line. For simplicity a static magnetic field is assumed, and the ambipolar potential is neglected. Then, outside the cyclotron resonance region the electron velocity is governed by the kinetic energy and magnetic moment conservation,

$$mc^2\gamma = \text{const}, \quad \gamma^2\beta_\perp^2/\omega_B = \text{const}, \quad (A1)$$

where conventional notations $\gamma = 1/\sqrt{1-\beta_\perp^2-\beta_\parallel^2}$, $\beta_{\perp,\parallel} = \upsilon_{\perp,\parallel}/c$, $\omega_B = eB/mc$ are used. Here particle movement along the field line corresponds to variation of the cyclotron frequency $\omega_B$. For definiteness we consider the electron velocity at the plane of "cold" cyclotron resonance where the cyclotron frequency is equal to the electromagnetic wave frequency, $\omega_B = \omega$. Further, all quantities calculated at this particular plane will be marked with primes, e.g. $\beta'_\perp$ and $\beta'_\parallel$ for the perpendicular and longitudinal electron velocities at cold resonance. The velocity at any other point is then

$$\beta_\perp = \sqrt{b}\beta'_\perp, \quad \beta_\parallel = \sqrt{\beta'^2_\parallel + \beta'^2_\perp(1-b)}$$

where $b = \omega_B/\omega = B/B'$.

The relativistic cyclotron resonance condition, $\omega(1-n_\parallel\beta_\parallel) = \omega_B/\gamma$, in our notation is

$$1-n_\parallel\sqrt{\beta'^2_\parallel + \beta'^2_\perp(1-b)} = b/\gamma'. \quad (A2)$$

Here $n_\parallel = k_\parallel c/\omega$ is the longitudinal index of refraction for the heating wave, and $\gamma = \gamma'$ (we use prime to denote that all velocities are defined at the cold resonance position). Formally, Eq. (A2) may be considered as a square equation for the resonant magnetic field $b$, which has real roots only if its determinant is positive:

$$(2-n_\parallel^2\gamma'\beta'^2_\perp)^2 - 4(1-n_\parallel^2(\beta'^2_\perp+\beta'^2_\parallel)) \geq 0.$$

This condition defines the region in a velocity space $(\beta'_\perp, \beta'_\parallel)$ for which the cyclotron resonance is possible at the fundamental harmonic. This condition may be simplified assuming $\beta'_\parallel \ll \beta'_\perp < 1$ (more exactly, neglecting $\beta'^4_\parallel$ and $\beta'^2_\parallel\beta'^2_\perp$ terms and retaining $\beta'^4_\perp$) as

$$\beta'^4_\perp \leq 4\beta'^2_\parallel/(2-n_\parallel^2). \quad (A3)$$

In other words, particles with large enough perpendicular velocities at the cold resonance can not reach the actual cyclotron resonance because they are reflected by the magnetic mirror.

The last effect can limit the maximum energy that particle can gain during multiple electron cyclotron interactions. Indeed, absorption of one "cyclotron photon" results in the variation of the particle energy and the perpendicular momentum as

$$\Delta mc^2\gamma = \hbar\omega, \quad \Delta p_\perp^2/2m = \hbar\omega_B.$$

First equation follows from the energy conservation for wave-particle interactions, second - from the Landau quantization. One can find that the momentum conservation, $\Delta p_\parallel = \hbar k_\parallel$, follows automatically from the above equations. Combining these equations one gets

$$\Delta(2\gamma/\omega - \gamma^2\beta_\perp^2/\omega_B) = 0.$$

Formally this equation is obtained at the actual resonance position. However, outside the narrow cyclotron interaction zone both terms in the l.h.s. are the invariants of motion (A1), so the equation holds at any reachable point along the field line. In particular, for the cold resonance cross-section, $\omega_B = \omega$, one obtains [37]

$$2\gamma' - \gamma'^2\beta'^2_\perp = \text{const}. \quad (A4)$$

This condition defines the direction of the particle flux induced by cyclotron waves in the velocity space (or called quasi-linear diffusion lines).

Let us consider a particle with some initial velocity $\beta'_{\|0}, \beta'_{\perp 0}$ specified at the cold resonance. Assuming that the particle gains large perpendicular velocity due to the cyclotron heating, $\beta'_{\|0} \approx \beta'_{\perp 0} \ll \beta'_\| \ll \beta'_\perp$, one may simplify Eq. (A4) as

$$\frac{1}{4}\beta'^4_\perp = \beta'^2_\| - \beta'^2_{\|0}. \tag{A5}$$

One can see that in zero order cyclotron heating of moderately hot (not ultra-relativistic) electrons results in increase of its gyrating energy ($\beta'_\perp$), while the energy of longitudinal motion is gained much slower ($\beta'_\| \sim \beta'^2_\perp$). Nevertheless, increase of parallel velocity is important for the existence condition of the cyclotron resonance given by Eq. (A3). Substituting $\beta'_\|$ from (A5) into (A3) one obtains [37]

$$\beta'^4_\perp \le 4\beta'^2_{\|0}/(1-n^2_\|). \tag{A6}$$

This limitation works only for $n^2_\| < 1$. So, in a rarified plasma there is a maximum electron energy that is defined by the initial velocity along the magnetic field. Note that the longitudinal velocity increase is small in this regime,

$$\beta'^2_\| \le \beta'^2_{\|0} \frac{2-n^2_\|}{1-n^2_\|}.$$

Let us find the position of the turning point (defined by $\beta_\| = 0$) corresponding to the maximum energy:

$$\frac{\delta B_{\text{turn}}}{B'} = b - 1 = \frac{\beta'^2_\|}{\beta'^2_\perp} = \beta'^2_\perp \times \frac{\beta'^2_\|}{\beta'^4_\perp} = \frac{2\beta'_{\|0}}{\sqrt{1-n^2_\|}} \times \frac{2-n^2_\|}{4} = \frac{1-n^2_\|/2}{\sqrt{1-n^2_\|}}\beta'_\|.$$

Here we use (A3) and (A6) considered as equalities. Evidently, the studied mechanism works until the turning point does not go outside the heating wave beam, or $\delta B_{\text{turn}} < \delta B_{\text{wave}}$, where $\delta B_{\text{wave}}$ is the variation of the magnetic field in the region highlighted by the wave beam.

In a dense but still collisionless plasma, $n^2_\| > 1$, the energy gain in the GDT conditions is limited by the finite size of the wave beam. Assuming $\max b = 1 + \delta b$, $\delta b \ll 1$, and $\beta'_\| \ll \beta'_\perp < 1$ in Eq. (A2), one gets

$$\delta b^2 + \delta b\, \beta'^2_\perp (n^2_\| - 1) = -\beta'^4_\perp/4 + n^2_\| \beta'^2_\|.$$

After substituting $\beta'_\|$ from Eq. (A5) and neglecting $\beta'_{\|0}$, one obtains the quadratic equation for $\alpha = \beta'^2_\perp / \delta b$,

$$(n^2_\| - 1)\alpha^2 - 4(n^2_\| - 1)\alpha - 4 = 0.$$

With variation of $\delta b \sim \delta B_{\text{wave}}/B'$ limited by the wave beam, the upper limit for the perpendicular velocity is defined as

$$\beta'^2_\perp < \alpha\, \delta b, \quad \alpha = 2 + \frac{2n_\|}{\sqrt{n^2_\| - 1}} \tag{A7}$$

where $\alpha$ is the largest root of the above quadratic equation. Note, that $\alpha \approx 4$ for $n_\| \gg 1$; neglecting the Doppler shift in the resonance condition would result in the two times lower limit, $\beta'^2_\perp \le 2\delta b$. Physically it means that escape of the cyclotron resonance from the wave beam spot occurs due to the relativistic mass, while the Doppler shift slows down such escape. Note that the final longitudinal velocity may be much greater than its initial value in this regime, $\beta'_\| \le \frac{1}{2}\alpha\delta b$.

Summarizing the results stated in Eqs. (A5)-(A7), the maximum electron energies at the cold resonance may be estimated as

$$E'_{\perp\max} \approx \begin{cases} \sqrt{\dfrac{2mc^2 E'_{\|0}}{1-n_\|^2}}, & n_\| < 1 \\ mc^2\left(1+\dfrac{n_\|}{\sqrt{n_\|^2-1}}\right)\dfrac{\delta B_{\text{wave}}}{B'}, & n_\| > 1 \end{cases}, \quad E'_{\|\max} \approx \begin{cases} E'_{\|0}\sqrt{\dfrac{2-n_\|^2}{1-n_\|^2}}, & n_\| < 1 \\ \dfrac{mc^2}{4}\left(1+\dfrac{n_\|}{\sqrt{n_\|^2-1}}\right)^2\left(\dfrac{\delta B_{\text{wave}}}{B'}\right)^2, & n_{\|2} > 1 \end{cases}.$$

These approximations both are not valid in the close vicinity of $n_\| = 1$, that corresponds to the so-called "auto-resonance" case [44, 45].

*A2. Solution of kinetic equation*

Under conditions (a)-(e) formulated in section 5.2, formation of the electron distribution function in phase space is governed by the Fokker–Plank kinetic equation. For simplicity we use the one dimensional distribution function over the total kinetic energy, $F(t,E)$, normalized over the total number of particles in a flux tube,

$$N(t) = \int_0^{E^*} F \, dE.$$

This distribution function may be found as a solution to the following kinetic equation [31]

$$\frac{\partial F}{\partial t} = \frac{\partial}{\partial E} ED \frac{\partial}{\partial E}\left(\frac{F}{\tau_b}\right) + J\delta(E), \tag{A8}$$

where

$$D(E) \approx 2\pi^2 |\text{Ai}(0)|^2 \left(\frac{e|E_-|}{mc\omega}\right)^2 \left(\frac{\omega L_B}{c}\right)^{4/3} \left(\frac{2mc^2}{E}\right)^{2/3} mc^2 \begin{cases} 1, & 0 < E < E^* \\ 0, & E > E^* \end{cases}$$

is the quasi-linear diffusion coefficient corresponding to the resonant field amplitude $E_-$; $\tau_b(E)$ is the half-period of the bounce oscillations defined earlier; the Dirac delta-function $J\delta(E)$ approximates the source of low-energy secondary electrons, in our model $J = N(t)\left(n_a \langle \upsilon \sigma_{\text{ion}} \rangle - \tau_h^{-1}\right)$. Details of derivation of the kinetic equation (A8) and limits of its applicability to mirror traps may be found in [31-35]. Note that Eq. (A8) is obtained for the particular point along the field line where $\omega_B = \omega$ ("cold resonance"), but in the present form it is applicable at any point since $E$ is an invariant of the adiabatic motion. The quasi-linear diffusion operator $D(E)$ describes here the cyclotron interaction in a non-uniform magnetic field in case when the ECR zone is close to the turning points of accelerating electrons, so electrons "remember" its phase between two consequent excursions through the ECR region [36].

The key parameter for the discussed theory is the time required for acceleration of a new-born electron up to the maximal energy $E_{\max}$,

$$\tau_{\text{ql}} \approx \tau_b E_{\max} / D(E^*).$$

Strictly speaking, the resonant field amplitude $E_-$ in this equation is a very uncertain parameter because the wave field is strongly varying in a resonant plasma. To take this into account, we define the wave field as the intensity averaged along the propagation ray:

$$\frac{c}{4\pi}|E_-|^2 = \langle I \rangle_{\text{ray}} = \frac{I_0}{L_{\text{ray}}} \int_0^{L_{\text{ray}}} \exp\left(-\int_0^l 2\,\text{Im}\, k\, dl'\right) dl.$$

The result is

$$\frac{c}{4\pi}|E_-|^2 \approx \frac{1-\exp(-\eta)}{\eta} \frac{1+\sin^2\theta}{2} \frac{P_{\text{gyr}}}{\pi a^2},$$

where $\eta$ is defined by Eq. (8), $\theta$ is the wave propagation angle to the magnetic field, and $P_{\text{gyr}}$ is the incident (gyrotron) power. The first term in the r.h.s. describes the field inhomogeneity, the second term projects the incident power to the left-hand circular polarization aligned to the external magnetic field, and the last term is the average density of the incident power flux. With this definition the acceleration time depends almost linearly on the plasma density inside the ECR zone. Figure A1 shows the result of calculation for the GDT parameters.

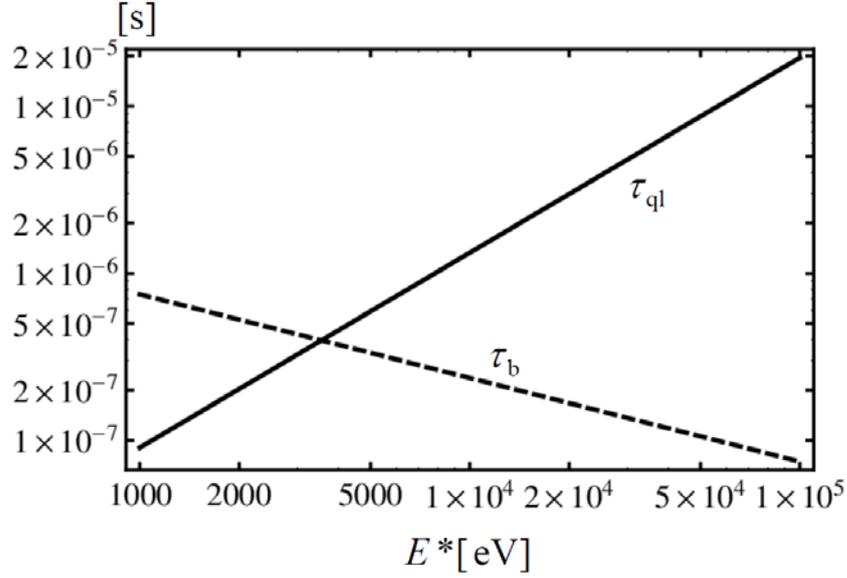

**Figure A1.** Acceleration time (solid line) and bounce-oscillation time (dashed line) v.s. maximum electron energy for the plasma density $\langle n_h \rangle_{ECR} = 2 \times 10^{11}$ cm$^{-3}$ and total microwave power 400 kW.

One can see that the acceleration time $\tau_{ql} \sim 10^{-6} - 10^{-5}$ s is much shorter then the plasma density growth time known from the experiment ($\sim 10^{-3}$ s). Under these conditions the distribution function forms the "plateau" in the region $0 < E < E^*$,

$$F(E)/\tau_b(E) = \text{const}. \tag{A9}$$

Normalizing this distribution over the total number of particles and using the bounce-oscillation time in the long trap limit, $\tau_b(E) \propto 1/\sqrt{E}$, one immediately obtains the solution of the kinetic equation given by Eq. (3).

*A3. Ionization rate*

Mean ionization rate corresponding to the plateau distribution may be calculated as

$$\langle v\sigma_{ion} \rangle_{pl} = N^{-1} \int_{E_{ion}}^{E^*} v\sigma_{ion} F_{pl}\, dE = \frac{1}{\sqrt{2mE^*}} \int_{E_{ion}}^{E^*} \sigma_{ion}\, dE.$$

The ionization cross-section may be taken in the following form [31, 46, 47]

$$\sigma_{ion} = \sigma_0 \begin{cases} \sqrt{2}\,\dfrac{x-1}{x(x+8)}, & 1 < x < 10 \\ \dfrac{1}{x}\ln\dfrac{x}{5}, & x > 10 \end{cases}$$

with $x = E/E_{ion}$. Averaging this expression one obtains

$$\langle v\sigma_{ion} \rangle_{pl} = \sigma_0 \frac{E_{ion}}{\sqrt{8mE^*}} \left( \sqrt{2}\ln\frac{4}{\sqrt[4]{5}} - \ln^2 2 + \ln^2 \frac{E^*}{5E_{ion}} \right).$$

Substituting $\sigma_0 \approx 1.4 \times 10^{-15}$ cm$^3$ and $E_{ion} \approx 15$ eV for deuterium, we obtain Eq. (4).

Figure A2 shows the dependence of the ionization rate on the limiting electron energy. For comparison, we plot here the ionization rate averaged over the Maxwellian distribution [48]. One can clearly see the effect of microwave heating—the significant increase of the ionization rate as compared to the thermal case with the same mean energy. Evidently, this increase is due to formation of a heavy electron tail in the electron distribution over energies. In our case, typical energies of hot electrons are in the range 5-10 keV. Corresponding increase of the ionization rate is about one order of magnitude, which is essential for interpretation of the experimental data.

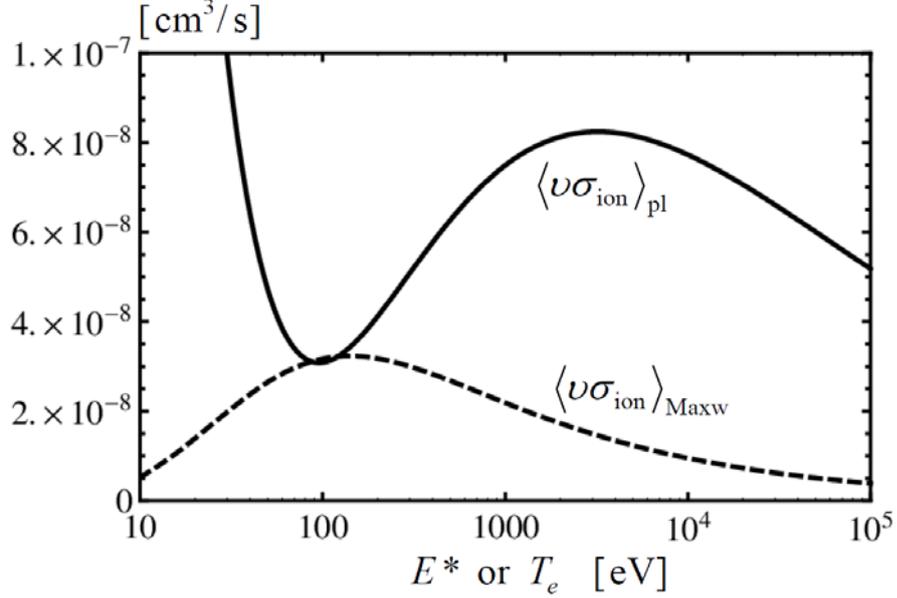

**Figure A2.** Ionization rate for plateau distribution v.s. maximum electron energy (solid line) and for Maxwellian distribution v.s. electron temperature (dashed line) for deuterium atoms.

*A4. Particle losses*

Here we estimate the confinement time $\tau_h$ of the accelerated electrons. In the regime of strong quasi-linear diffusion, $\tau_{ql} \ll \tau_h$, the loss-cone in the momentum space is fully filled with particles driven by RF field. As a good approximation one can assume that the quasi-linear plateau is formed in the whole two-dimensional momentum space including the loss-cone region. Then, the loss rate of the hot electrons can be calculated as particle flux into the loss-cone for the known distribution function [31, 37]. The same result may be expressed in terms of one dimensional distribution $F(E)$ assuming that corresponding two-dimensional distribution is strongly anisotropic in the resonance region, $\langle E_\perp \rangle \gg \langle E_\parallel \rangle$, where $\langle ... \rangle$ denotes the mean value at the ECR surface. Then, the loss-cone is defined by condition

$$E \approx E_\perp < E_{lc} = \frac{\langle E_\parallel \rangle}{R'_{ECR} - 1} \ll \langle E \rangle,$$

where $R'_{ECR} = B_{plug} / B_{ECR}$ is the "external" mirror ratio of the ECR zone, $B_{plug}$ and $B_{ECR}$ are the magnetic fields at the mirror plugs and ECR zone. The average life-time of the particle inside the loss-cone may be defined by the half-period of the bounce oscillations

$$\tau_b(E) \approx \frac{L}{\sqrt{2E/m}\sqrt{1 - 1/R_{ECR}}}$$

calculated with taking into account the strong anisotropy of the electron distribution function for a long trap. Here $L$ is the mirror-to-mirror length; $R_{ECR} = B_{ECR} / B_{center}$ is the "internal" mirror ratio of the ECR zone. The loss rate may be defined as

$$\frac{1}{\tau_h} \equiv \left\langle \frac{1}{\tau_b} \right\rangle_{pl} = N^{-1} \int_0^{E_{lc}} \frac{F_{pl}(E)}{\tau_b(E)} dE \approx N^{-1} \frac{\langle E_\parallel \rangle}{R'_{ECR} - 1} \lim_{E \to 0} \frac{F_{pl}(E)}{\tau_b(E)}.$$

Here we take into account Eq. (A9). Substituting the plateau distribution function (3) we obtain the loss rate of the hot electrons given by Eq. (5).

*A5. Energy losses*

Direct energy losses of hot electrons associated with the flux into the loss cone may be calculated in a similar way as the particle losses

$$\left\langle \frac{E}{\tau_b} \right\rangle_{pl} = N^{-1} \int_0^{E_{lc}} E \frac{F_{pl}(E)}{\tau_b(E)} dE \approx \frac{E_{lc}}{2\tau_h}.$$

These losses are typically negligible because $E_{lc}$ is much less then the mean energy of hot electrons $\langle E \rangle_{pl} = \frac{1}{3} E^*$, see Eq. (7).

Energy losses due to collisional friction of hot electrons with the bulk electrons may be estimated as

$$n_h \langle \nu_{eh} E \rangle_{pl} = n_h \nu_{eh}^* \int_{T_e}^{E^*} \nu_{eh}(E) E F_{pl} dE = n_h \nu_{eh}^* E^* \ln \sqrt{\frac{E^*}{T_e}},$$

where $\nu_{eh}^* = \nu_{eh}(E^*)$, $\nu_{eh}(E) \approx 7.7 \times 10^{-6} \Lambda n_e E^{-3/2}$ is the Coulomb energy transfer rate from energetic to cold electrons, $\nu_{eh}$ is in 1/s, $n_e$ is in cm$^{-3}$, $E$ is in eV, and $\Lambda \approx 12$ is Coulomb logarithm [49]. Note that these losses are primarily due to electrons with energies much lower then $E^*$; formally it is described by large logarithmic factor $\ln\sqrt{E^*/T_e} \sim 3$.

To obtain the energy losses that come to the secondary electrons we need its distribution function over energies. For that, we adopt the approximation for the double ionization cross-section from Ref. [50]:

$$\sigma_2(E, E') = \frac{C(E)}{1 + (E'/E_0)^2} \begin{cases} 1, & E' \leq (E - E_{ion})/2 \\ 0, & E' > (E - E_{ion})/2 \end{cases}.$$

Here $E$ and $E'$ are the energies of the primary and secondary electrons, correspondingly, $E_0$ is the constant obtained by fitting of the experimental data (for hydrogen $E_0 = 8.3$ eV, so we assume the same value for deuterium), and the function

$$C(E) = \frac{\sigma_{ion}(E)}{E_0 \arctan\frac{E - E_{ion}}{2E_0}}$$

is obtained from the condition $\int_0^{(E-E_{ion})/2} \sigma_2(E, E') dE' = \sigma_{ion}(E)$. The net power that goes to the secondary electrons is then calculated as

$$n_h n_a \langle \upsilon \sigma_2 E' \rangle_{pl} = n_h n_a N^{-1} \int_{E_{ion}}^{E^*} \int_0^{(E-E_{ion})/2} \upsilon \sigma_2 E' F_{pl} \, dE' dE =$$

$$= n_h n_a \frac{E_0^2}{\sqrt{2mE^*}} \int_{E_{ion}}^{E^*} C(E) \ln\sqrt{1 + (E - E_{ion})^2/4E_0^2} \, dE \approx$$

$$\approx n_h n_a \frac{\sigma_0 E_{ion} E_0}{\pi\sqrt{2mE^*}} \ln^2 \frac{E^*}{E_{ion}} \cdot \ln \frac{(E^*)^{2/3} E_{ion}^{1/3}}{10 E_0}.$$

The result may be conveniently presented as the effective mean energy of the secondary electrons,

$$E_{sec} = \langle \upsilon \sigma_2 E' \rangle_{pl} / \langle \upsilon \sigma_{ion} \rangle_{pl}.$$

One can find that $E_{sec}$ is of the order of the ionization energy if $E^* \gg E_{ion}$. In deuterium plasma $E_{sec}$ is slowly increasing from 16 eV at $E^* = 1$ keV to 30 eV at $E^* = 100$ keV. For the GDT conditions $E^* = 6.5$ keV and $E_{sec} = 23$ eV.